\documentclass[twocolumn,showpacs,preprintnumbers,amsmath,amssymb,superscriptaddress,floatfix]{revtex4}

\usepackage{graphicx}
\usepackage{dcolumn}
\usepackage{bm}
\usepackage{dcolumn}
\usepackage[]{natbib}

\begin{document}

\preprint{SKC/002-ANU}

\title{Measured $g$~factors and the tidal-wave
description of transitional nuclei near $A = 100$}

\author{S.K. Chamoli}
\affiliation{Department of Nuclear Physics, Research School of
Physics and Engineering, Australian National University, Canberra,
ACT 0200, Australia}

\author{A.E. Stuchbery}
\affiliation{Department of Nuclear Physics, Research School of
Physics and Engineering, Australian National University, Canberra,
ACT 0200, Australia}

\author{S. Frauendorf}
\affiliation{Department of Physics, University of Notre Dame, Notre
Dame, IN 46556, USA}

\author{J. Sun}
\affiliation{Department of Physics, University of Notre Dame, Notre
Dame, IN 46556, USA}

\author{Y. Gu}
\affiliation{Department of Physics, University of Notre Dame, Notre
Dame, IN 46556, USA}

\author{R.F. Leslie}
\affiliation{Department of Nuclear Physics, Research School of
Physics and Engineering, Australian National University, Canberra,
ACT 0200, Australia}

\author{P.T. Moore}
\affiliation{Department of Nuclear Physics, Research School of
Physics and Engineering, Australian National University, Canberra,
ACT 0200, Australia}

\author{A. Wakhle}
\affiliation{Department of Nuclear Physics, Research School of
Physics and Engineering, Australian National University, Canberra,
ACT 0200, Australia}

\author{M.C. East}
\affiliation{Department of Nuclear Physics, Research School of
Physics and Engineering, Australian National University, Canberra,
ACT 0200, Australia}

\author{T. Kib\'edi}
\affiliation{Department of Nuclear Physics, Research School of
Physics and Engineering, Australian National University, Canberra,
ACT 0200, Australia}

\author{A.N. Wilson}
\affiliation{Department of Nuclear Physics, Research School of
Physics and Engineering, Australian National University, Canberra,
ACT 0200, Australia}


\date{\today}

\begin{abstract}
The transient-field technique has been used in both conventional
kinematics and inverse kinematics to measure the $g$~factors of the
2$^+_1$ states in the stable even isotopes of Ru, Pd and Cd. The
statistical precision of the $g(2^+_1)$ values has been
significantly improved, allowing a critical comparison with
the tidal-wave version of the cranking model recently proposed for
transitional nuclei in this region.
\end{abstract}


\maketitle

\section{Introduction}

The stable isotopes of $_{42}$Mo, $_{44}$Ru, $_{46}$Pd, and $_{48}$Cd, include some of the best examples of vibrational level structures, with $^{110-116}$Cd, in particular, frequently being cited as `textbook' examples \cite{BMvol2p532,IAp42}. Recent studies indicate that the vibrational picture is reasonably good at the two-phonon level in $^{110-114}$Cd, but breaks down at the three-phonon level, particularly for non-yrast states \cite{garrett08}.

The lower mass stable isotopes and neutron-deficient isotopes of these elements are near $^{90}_{40}$Zr$_{50}$, which is almost double magic, and $^{100}_{\;\: 50}$Sn$_{50}$, which is double magic. Having few valence nucleons, the level schemes therefore show spherical structures and are accessible to shell model calculations \cite{Holt07}. By way of contrast with the spherical and vibrational structures in the low and intermediate mass isotopes, the heavier isotopes, with neutron numbers near midshell, make a transition to rotational structures. There has been considerable effort in recent years to study the spectroscopy of isotopes in this region produced either as fission fragments or as radioactive beams. Examples of experimental work relevant to the present study are measurements of quadrupole moments and $B(E2)$ values in neutron-deficient Cd isotopes produced as radioactive beams \cite{ekstrom09}, and measurements of $g$~factors in neutron-rich fission fragments, in which a reduced magnitude for several neutron-rich nuclei was attributed to contributions from neutrons in the $h_{11/2}$ orbit \cite{smith04,smith05}.

On the experimental side, the present work focuses on measurements of the $g$~factors of the first excited states in all of the stable even isotopes of Ru, Pd and Cd by the transient-field technique. The precision is improved considerably compared with previous work.

On the theoretical side, we use the tidal wave approach for calculating the $g$~factors in this transitional region. The model uses the fact that in semi-classical approximation the yrast states of vibrational nuclei correspond to quadrupole waves traveling over the surface of the nucleus like the tidal waves over the surface of the ocean. It has been demonstrated that the energies of the yrast states, as well as the $B(E2)$ values of the transitions between them, are very well described by this model for the even-even nuclei with $44\leq Z \leq 48$ and $54\leq N \leq 68$~ \cite{frauendgusun1,frauendgusun2}. The present work extends the model to $g$~factors, which allows an examination of the way in which the angular momentum is shared between the protons and neutrons.

The paper is arranged as follows: Section \ref{sect:experiment} reports the $g$-factor experiments. The measurements using conventional kinematics are described first (sect. \ref{sect:conv}), followed by the measurements using inverse kinematics (sect. \ref{sect:inv}). A summary and discussion of adopted experimental $g$~factors in sect. \ref{sect:adopted} completes the experimental part of the paper. The tidal wave model calculations of the $g$~factors are presented in sect. \ref{sect:theory} and the comparison between theory and experiment is discussed in sect. \ref{sect:discussion}. The conclusion follows (sect. \ref{sect:conclusion}).

\section{Transient-field $g$-factor measurements}
\label{sect:experiment}

The $g$~factors of the first excited states were measured in all the
stable even isotopes of Ru, Pd and Cd using the transient-field
technique and beams from the Australian National University 14UD
Pelletron accelerator. Measurements on the Cd and Ru isotopes in
`conventional kinematics' are described in sect. \ref{sect:conv};
those on the Ru, Pd and Cd isotopes in `inverse kinematics' are
described in sect.~\ref{sect:inv}. The experiments used the ANU
Hyperfine Spectrometer \cite{hyperion}. Experimental
procedures were similar to those described elsewhere
\cite{stu85,Man2001,spe02,ben07,stu07,eas09,eas09a}.

Before describing the experiments we review some procedures and terminology associated with the determination of the experimental $g$~factors from transient-field precession measurements \cite{ben80,stu85,spe02,ben07}.

The observed transient-field precession, $\Delta \Theta_{\rm obs}$, is related to the nuclear $g$~factor, $g$, by
\begin{equation} \label{eq:DthObs}
\Delta \Theta_{\rm obs} = g \phi(\tau)
\end{equation}
where
\begin{equation} \label{eq:phiTF}
\phi(\tau)=-\frac{\mu_N}{\hbar} \int_{t_i}^{t_e} B_{\rm TF}(v(t),Z) e^{-t/\tau} dt,
\end{equation}
and $\mu_N$ is the nuclear magneton; $\tau$ is the mean life of the nuclear state. The transient field strength $B_{\rm TF}(v(t),Z)$ depends on the atomic number and velocity of the ion within the ferromagnetic layer of the target. It is often parametrized in the form
\begin{equation} \label{eq:Bparam}
 B_{\rm TF}(v,Z) = a_{\rm TF} Z^{p_Z} (v/v_0)^{p_v}.
 \end{equation}
For fully magnetized iron hosts the Rutgers parametrization gives $a_{\rm TF} = 16.9$~T,  ${p_Z} = 1.1$, and $p_v = 0.45$ \cite{shu80}.

As can be seen from Eq.~(\ref{eq:phiTF}), $\phi$ is a function of $\tau$. In the present work $\tau > 4$~ps while $t_e$ is typically about 0.5 ps, so the exponential factor in Eq.~(\ref{eq:phiTF}) remains near unity; the observed precession is insensitive to $\tau$, but not independent of it, especially for the shorter-lived states. Furthermore, for each isotope the reaction kinematics, and slowing down of the ions in the ferromagnetic layer, are slightly different. It is useful to define the limiting case of $\phi$ for $\tau \rightarrow \infty$, namely
\begin{equation} \label{eq:phi-infty}
\phi(\infty) =-\frac{\mu_N}{\hbar} \int_{t_i}^{t_e} B_{\rm TF}(v(t),Z) dt.
\end{equation}

In the following presentation of experimental results, the observed precession angles $\Delta \Theta_{\rm obs}$ will always be given. Depending on what is known about the transient-field strength for the particular combination of ion and ferromagnetic host, it may be useful to present, in addition, a set of corrected experimental precessions for each isotope $A$, which reflect the relative $g$~factors:
\begin{equation} \label{eq:dth-corr}
\Delta \Theta(A) =\Delta \Theta_{\rm obs}(A) \phi(\infty, A_{\rm ref})/\phi(\tau, A),
\end{equation}
where $A_{\rm ref}$ denotes a chosen reference nuclide. The ratio of calculated $\phi$ values effectively removes the small differences in the measured precession angles due to differences in level lifetimes and reaction kinematics. In the present work $\phi(\infty, A_{\rm ref})/\phi(\tau, A)$ is near unity. It is independent of the chosen transient field scale parameter $a_{\rm TF}$, and is also insensitive to any reasonable choice of $p_Z$, and $p_v$.

In some cases it is appropriate to give experimental $g$~factors relative to a reference $g$~factor in the nucleus $A_{\rm ref}$, namely $g_{\rm ref}(A_{\rm ref})$:
\begin{equation} \label{eq:gA-gref}
g(A) = g_{\rm ref}(A_{\rm ref}) \frac{\Delta \Theta_{\rm obs}(A)}{\Delta \Theta_{\rm obs}(A_{\rm ref})}
\frac{\phi(\tau_{\rm ref}, A_{\rm ref})}{\phi(\tau, A)},
\end{equation}
where $\tau_{\rm ref}$ is the mean life of the reference state.

\subsection{Conventional kinematics} \label{sect:conv}

\subsubsection{Cd isotopes} \label{sect:convCd}

The $g$~factors of the first $2^+_1$ states in
$^{110,112,114,116}$Cd were measured simultaneously, relative to
each other, using the transient-field technique in conventional
kinematics. The experiment was similar to that on the Mo isotopes
\cite{Man2001}. Table \ref{tab:Cdkinem} summarizes the relevant
level properties and reaction kinematics.

\begin{table*}
\caption{Level properties and reaction kinematics for Cd and Ru isotopes recoiling
in iron after Coulomb excitation by 95 MeV $^{32}$S beams.
$E(2_1^+)$ is the energy and $\tau(2^+_1)$ is the mean life of the
2$^+_1$ level \protect \cite{Raman}. $\langle E_i \rangle$ and $\langle E_e \rangle$ are
the average energies with which the Cd ions enter into and exit from
the iron foil. The corresponding ion velocities are $\langle v_i/v_0
\rangle$ and $\langle v_e/v_0 \rangle$. The average ion velocity is
$\langle v/v_0 \rangle$. $v_0 = c/137$ is the Bohr velocity.
$t_{\rm Fe}$ is the effective time for which the ions experience the transient field
in the iron layer of the target. These
quantities were calculated with the stopping powers of Ziegler {\it
et al.} \protect\cite{zie85}. $\phi(\tau)$ is the transient-field
precession per unit $g$~factor calculated as described in the text.}
\label{tab:Cdkinem}
\begin{ruledtabular}
\begin{tabular}{cccccccccc}
Isotope & $E(2^+_1)$ & $\tau(2^+_1)$ & $\langle E_i \rangle$ &
$\langle E_e \rangle$ & $\langle v_i/v_0 \rangle$ & $\langle v_e/v_0
\rangle$ & $\langle v/v_0 \rangle$ & $t_{\rm Fe}$ &
$\phi(\tau)$ \\
 & (keV) & (ps) & (MeV) & (MeV) & & & & (fs) & (mrad)\\ \hline
$^{110}$Cd & 657 &  7.4 & 51.6 & 8.4 & 4.35 & 1.75 & 2.77 & 543 & $-48.8$  \\
$^{112}$Cd & 617 &  8.9 & 51.1 & 8.5 & 4.29 & 1.75 & 2.75 & 552 & $-49.4$  \\
$^{114}$Cd & 559 & 13.7 & 50.7 & 8.6 & 4.23 & 1.75 & 2.72 & 564 & $-50.2$  \\
$^{116}$Cd & 512 & 20.3 & 50.3 & 8.7 & 4.18 & 1.74 & 2.70 & 572 & $-50.8$  \\
 &\\
$^{100}$Ru & 540 & 18.2 & 58.6 & 12.3 & 4.86 & 2.23 & 3.34 & 410 & $-36.5$  \\
$^{102}$Ru & 475 & 26.6 & 58.1 & 12.1 & 4.79 & 2.21 & 3.30 & 417 & $-36.2$  \\
$^{104}$Ru & 358 & 83.4 & 57.6 & 12.5 & 4.72 & 2.20 & 3.26 & 426 & $-37.3$
 \end{tabular}
\end{ruledtabular}
\end{table*}

States of interest were Coulomb excited using beams of 95 MeV
$^{32}$S.  In the order encountered by the beam, the target
consisted of layers of $^{\rm nat}$Ag, 0.05 mg/cm$^2$ thick, and
$^{\rm nat}$Cd, 0.98 mg/cm$^2$ thick, which had been evaporated onto
an annealed iron foil, 2.64 mg/cm$^2$ thick. On the back of the iron
foil a 5.47 mg/cm$^2$ thick layer of natural copper had already been
evaporated. For additional mechanical support, and improved thermal
contact with the cooled target mount, this multilayered target was
pressed onto thicker ($\sim$ 12 $\mu$m) copper foil using an
evaporated layer of indium as adhesive. Coulomb excited Cd nuclei
recoiled through the iron foil, where they experienced the transient
field, and then stopped in the non-magnetic copper layer where they
subsequently decayed. The thin Ag layer on the front of the target
was included to help protect the Cd layer, which has a low melting
point of $\sim 321^{\circ}$C. To minimize the effect of beam heating
on the target, it was maintained at a temperature of 6K by mounting
it on a cryocooler (Sumitomo RDK-408D). No deterioration of the Cd
target layer was observed despite a high beam current of $\sim 12.5$
pnA being maintained throughout the measurement ($\sim 4$ days).

An external magnetic field of 0.09~T was applied perpendicular to
the $\gamma$-ray detector plane to magnetize the ferromagnetic layer
of the target. The direction of this field was reversed periodically
to minimize systematic errors.

\begin{figure*}[t]
  \includegraphics[width=0.90\textwidth]{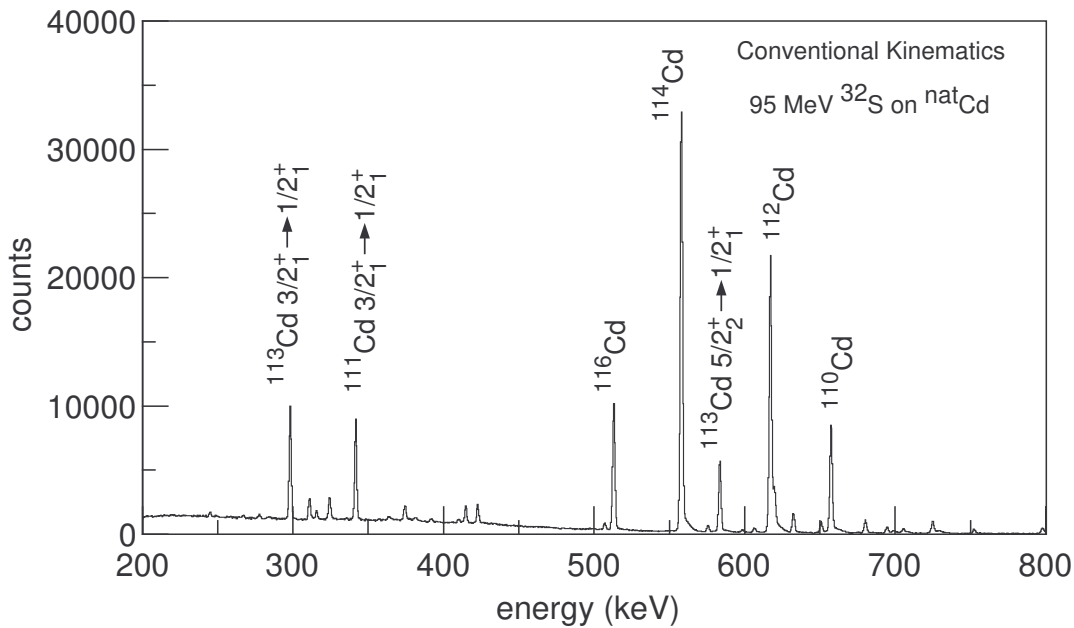}
  \caption{Spectrum of $\gamma$-rays observed at $+65^\circ$ to the
beam axis in coincidence with backscattered $^{32}$S beam ions. This spectrum
represents all of the data for the field up direction in the
detector at $+65^\circ$, obtained during the precession
measurement on the Cd target.}
  \label{fig:Cdspectrum}
\end{figure*}

Backscattered beam ions were detected in a pair of silicon
photodiode detectors, 10 mm high by 9 mm wide, placed 3.8 mm
from the beam axis in the vertical plane parallel to the target, and 16
mm upstream of the target; the average scattering angle was
151$^{\circ}$. To measure the transient-field precession, $\gamma$
rays emitted in coincidence with backscattered particles were
observed in two 50\% (relative efficiency) HPGe detectors and two
20\% HPGe detectors placed at $\pm 65^\circ$ and $\pm 115^\circ$ to
the beam axis, respectively. The target-detector distances were set
so that the detector crystals all subtended a half angle of
$18^\circ$. Figure \ref{fig:Cdspectrum} shows a coincidence
$\gamma$-ray spectrum observed in the detector at $+65^\circ$ to the
beam direction.

\begin{figure}[t]
  \includegraphics[width=0.48\textwidth]{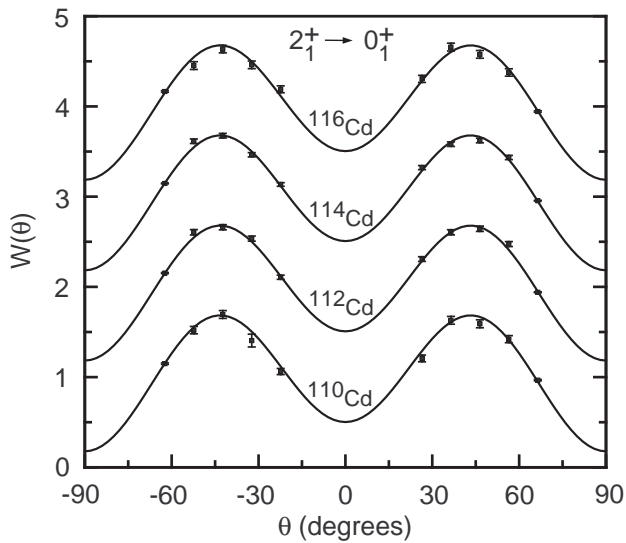}
\caption{Experimental and calculated angular correlations for
even-$A$ Cd isotopes following Coulomb excitation with 95 MeV
$^{32}$S beams. (Data for the different isotopes have been offset
for presentation.)}
  \label{fig:CdAD}
\end{figure}

Particle-$\gamma$ angular correlations were measured in a sequence
of runs of about 75 min duration. The backward-placed Ge detectors
were kept at $\pm 115^\circ$, to normalize the number of counts,
while the angular correlation was sampled with the two forward Ge
detectors at $\pm 25^\circ$, $\pm 35^\circ$, $\pm 45^\circ$, $\pm
55^\circ$ and $\pm 65^\circ$, in turn. The measured angular
correlations for the $2^+_1 \rightarrow 0^+_1$ transitions are compared with the calculated angular correlations
in Fig.~\ref{fig:CdAD}.

The transient-field precession angles, $\Delta \Theta$, were
determined by the usual procedures \cite{ben80,stu85,rob99,spe02}.
Briefly, $\Delta \Theta = \epsilon /S$, where $\epsilon$ is the
`effect' and $S(\theta_{\gamma})=(1/W){dW}/{d\theta}$, often referred to
as the `slope', is the logarithmic derivative of the angular correlation at the
$\gamma$-ray detection angle, $\theta_{\gamma}$. Formally, $\epsilon = (N\downarrow -
N\uparrow)/(N\downarrow + N\uparrow)$, where $N\uparrow
(\downarrow)$ refers to the counts recorded for field up (down) at
$+\theta_\gamma$; however the evaluation of $\epsilon$ from the
experimental data proceeds via the formation of a double ratio of
counts recorded for field up and down in a pair of detectors at $\pm
\theta_{\gamma}$.

\begin{table*}
\caption{Results of precession measurements and $g$~factors for the
2$^+_1$ states in Cd isotopes (conventional kinematics). $|S(65^\circ)|$ is the logarithmic
derivative of the angular correlation. $ \Delta \Theta _F$ and $
\Delta \Theta_B$ are the precession angles observed with the
$\gamma$-ray detector pairs at $\pm 65^\circ$ and $\pm 115^\circ$,
respectively; $\langle \Delta \Theta \rangle_{\rm obs} $ is the average of
these. $ \Delta \Theta $ has been corrected for small
differences in the reaction kinematics as the ions traverse the iron
layer of the target, so that relative values of $ \Delta \Theta $ give relative $g$~factors.} \label{tab:Cdresults}
\begin{ruledtabular}
\begin{tabular}{ccccccll}
Isotope & $|S(65^\circ)|$ & $\Delta \Theta _F$ & $ \Delta \Theta_B$
&
$\langle  \Delta \Theta \rangle_{\rm obs} $  & $ \Delta \Theta$  & $g/g(^{116}{\rm Cd})$  & $g$ \footnotemark[1]\\
 & (rad$^{-1}$) & (mrad) & (mrad) &  (mrad) &  (mrad) &\\
\hline $^{106,108}$Cd \footnotemark[2]
              & 2.8  & $-$18(4)   & $-$27(4)   & $-$21.7(26) &$-$22.2(27) & 1.51(20) & +0.44(6)  \\
$^{110}$Cd    & 2.74 & $-$19.9(9) & $-$19.7(12)& $-$19.79(74)&$-$20.32(76)& 1.38(9)  & +0.404(29) \\
$^{112}$Cd    & 2.71 & $-$17.3(6) & $-$18.2(8) & $-$17.61(48)&$-$17.90(49)& 1.22(7)  & +0.356(24) \\
$^{114}$Cd    & 2.69 & $-$15.9(5) & $-$16.6(6) & $-$16.18(39)&$-$16.18(39)& 1.10(6)  & +0.321(21) \\
$^{116}$Cd    & 2.70 & $-$14.9(10)& $-$14.7(13)& $-$14.87(78)&$-$14.73(77)& 1        & +0.292(24) \\
\end{tabular}
\footnotetext[1]{~Assigned errors include the uncertainty in the transient-field calibration.}
\footnotetext[2]{~Results for the composite 635 keV line, which
includes both $^{106}$Cd and $^{108}$Cd, are included to show
consistency with the measurements on these isotopes reported below.}
\end{ruledtabular}
\end{table*}

Results of the precession measurements on the 2$^+_1$ states of the
even Cd isotopes are given in Table \ref{tab:Cdresults}. Differences
in the `slopes', $S(65^\circ)$, stem mainly from differences in the
small level of feeding intensity from the higher excited states,
especially the 4$^+_1$ state. The effect of this feeding contribution on the extracted $g$~factors was evaluated as described in Ref.~\cite{stu07}. It was found
that {\em extreme} values must be assumed for the magnitude of
$g(4^+_1)$ in order to make even a few percent change in the
precession observed for the 2$^+_1$ state. The effect of the feeding
contribution is therefore accurately included in the present
analysis by evaluating $S$ for the fed (i.e. observed) angular
correlation for the 2$^+_1$ state.

The absolute values of the $g$~factors in the Cd isotopes were
determined by reference to $^{106}$Pd in the inverse kinematics
measurements described below. Specifically, the precessions measured (in inverse kinematics)
for $^{112}$Cd and $^{114}$Cd beams, gave $g(^{112}{\rm
Cd})=+0.365(30)$ and $g(^{114}{\rm Cd})=+0.313(25)$. Combining these
values with the precessions for these isotopes in
Table~\ref{tab:Cdresults} determines that the absolute $g$~factors
in Table~\ref{tab:Cdresults} are given by $g =
\Delta \Theta/\phi(\infty) = \Delta \Theta/(-50.34 \pm 3.08)$, where $\Delta \Theta$ is in mrad.

A measurement on $^{110}$Cd by Benczer-Koller {\em et al.} \cite{ben89}, in which the transient field strength was determined by the Rutgers parametrization \cite{shu80}, gave $g(2^+_1;^{110}{\rm Cd})= +0.382(17)$, in very good agreement with the value obtained
here. Our work therefore confirms the Rutgers parametrization of the
transient-field strength for ions with $Z \sim 46 - 48$ traversing
iron foils with velocities in the range $2v_0 \lesssim v \lesssim
4v_0$. For this reason, the values  of $\phi(\tau)$ shown
in Table~\ref{tab:Cdkinem} were evaluated using the Rutgers parametrization
\cite{shu80} in Eq.~(\ref{eq:phiTF}).

The $g$~factors of several states in the two odd-$A$ isotopes, $^{111}$Cd and
$^{113}$Cd, were measured as a by-product of the
measurement on the natural target. These results will be presented and discussed elsewhere \cite{oddACd}.

\subsubsection{Ru isotopes} \label{sect:convRu}

The $g$~factors of the first $2^+_1$ states in
$^{100,102,104}$Ru were measured simultaneously, relative to
each other, using the transient-field technique in conventional
kinematics and procedures very similar to those described in sect.~\ref{sect:convCd}. States of interest were again Coulomb excited using beams of 95 MeV $^{32}$S.  The target
consisted of a layer of $^{\rm nat}$Ru, 0.63 mg/cm$^2$ thick, which had been sputtered onto an annealed iron foil, 2.34 mg/cm$^2$ thick. The iron foil was then pressed onto a 12.5 $\mu$m thick copper foil using an evaporated layer of indium, $\sim 3 $~mg/cm$^2$ thick, as adhesive. Coulomb excited Ru nuclei recoiled through the iron foil, where they experienced the transient field, and then stopped in the non-magnetic indium and copper layers where they subsequently decayed. (Both indium and copper have cubic crystalline structure so quadrupole interactions for the 2$^+$ states of the Ru isotopes are negligible in both host materials.)

\begin{figure*}[t]
  \includegraphics[width=0.90\textwidth]{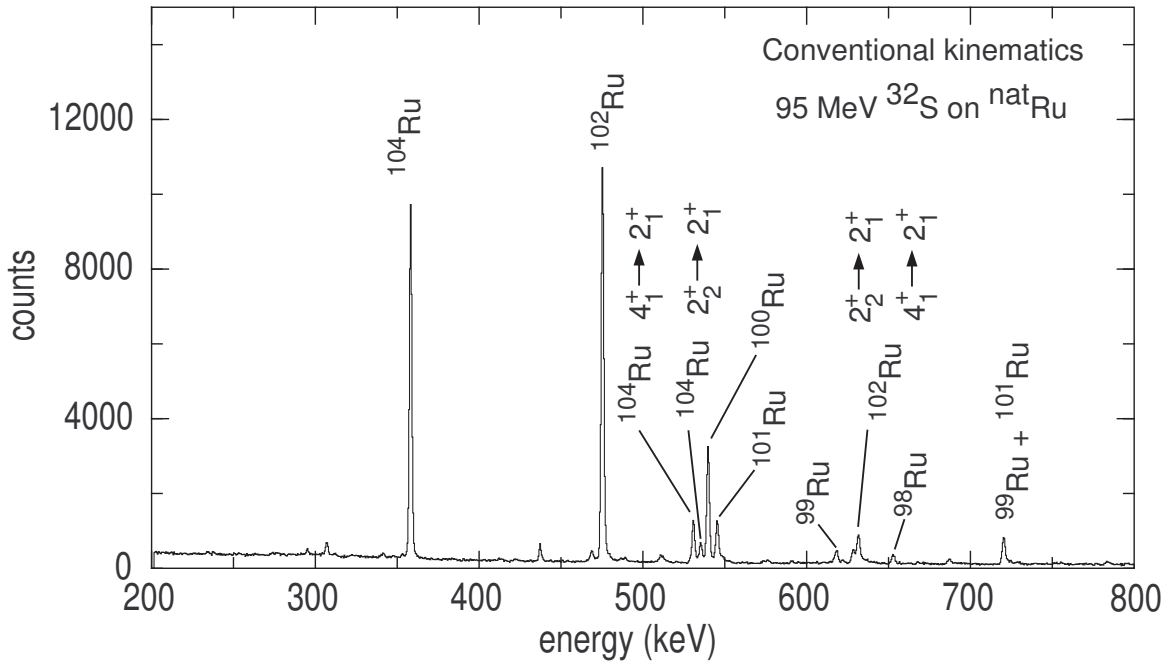}
  \caption{Spectrum of $\gamma$-rays observed at $+65^\circ$ to the
beam axis in coincidence with backscattered $^{32}$S beam ions. This spectrum
represents the data for the field up direction in the
detector at $+65^\circ$, obtained during the precession
measurement on the natural Ru target.}
  \label{fig:Ruspectrum}
\end{figure*}

Save for the different target, the experiment was essentially identical to that on the Cd isotopes reported in the previous section. The total beam time for the precession measurement was about 60 hours. Table \ref{tab:Cdkinem} includes a summary of the relevant
level properties and reaction kinematics. Figure \ref{fig:Ruspectrum} shows a coincidence $\gamma$-ray spectrum observed in the detector at $+65^\circ$ to the beam direction. Although natural Ru contains $^{96}$Ru (5.5\%), $^{98}$Ru (1.9\%), $^{99}$Ru (12.7\%), and
$^{101}$Ru (17.0\%), along with  $^{100}$Ru (12.6\%), $^{102}$Ru (31.6\%) and $^{104}$Ru (18.7\%), transient-field precessions could be obtained with meaningful precision only for the latter three isotopes.

\begin{figure}[t]
  \includegraphics[width=0.48\textwidth]{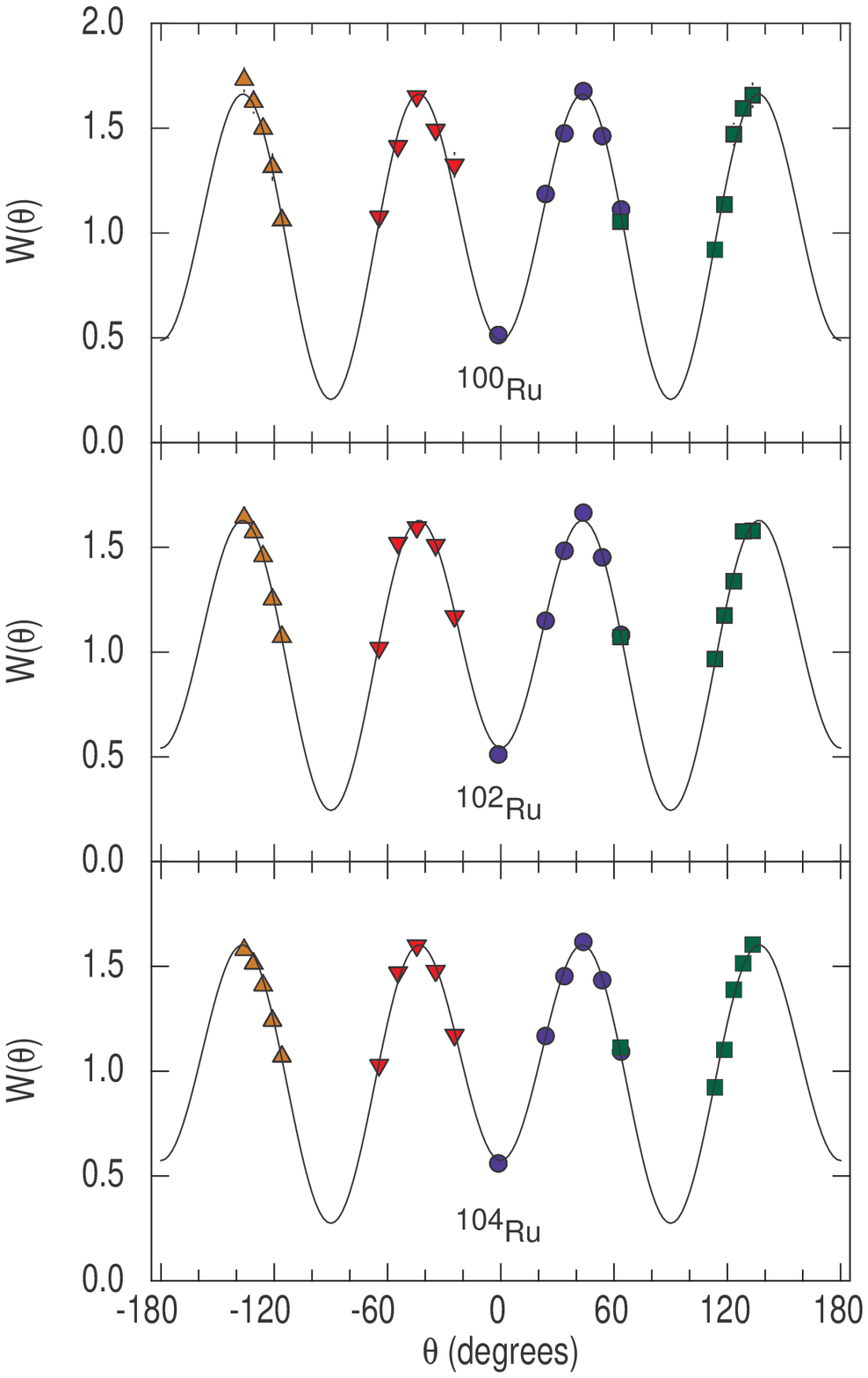}
\caption{(Color online) Experimental and calculated angular correlations for
even-$A$ Ru isotopes following Coulomb excitation with 95 MeV
$^{32}$S beams. Different symbols represent the four $\gamma$-ray detectors.}
  \label{fig:RuAD}
\end{figure}

Particle-$\gamma$ angular correlations were measured in a sequence
of runs of about 50 min duration. The two Ge detectors at negative angles were kept at $- 65^\circ$ and $- 115^\circ$, to normalize the number of counts, while the angular correlation was sampled at a sequence of angles in the positive hemisphere with the other two Ge detectors. The procedure was then reversed: the detectors at positive angles remained fixed while the detectors at negative angles were moved to a sequence of angles. The measured angular correlations are compared with the calculated angular correlations in Fig.~\ref{fig:RuAD}.

\begin{table}
\caption{Results of precession measurements and $g$~factors for the
2$^+_1$ states in Ru isotopes (conventional kinematics). The transient-field strength was calibrated using the Rutgers parametrization \protect\cite{shu80}.} \label{tab:Ruresults}
\begin{ruledtabular}
\begin{tabular}{ccclc}
Isotope & $|S(65^\circ)|$ & $\langle  \Delta \Theta \rangle $  &$g/g(^{104}{\rm Ru})$ &   $g$ \footnotemark[1]\\
 &  (rad$^{-1}$) &  (mrad) &\\
\hline
$^{100}$Ru    & 2.72 & $-$15.88(84)& 1.08(7) & +0.434(23) \\

$^{102}$Ru    & 2.59 & $-$16.85(44)& 1.14(5) & +0.457(12) \\

$^{104}$Ru    & 2.50 & $-$14.94(45)& 1       & +0.401(12) \\
\end{tabular}
\footnotetext[1]{~No uncertainty in the transient-field calibration is included here. Adopted $g$~factors, including uncertainties in the field strength, are presented in sect.~\ref{sect:adopted}.}
\end{ruledtabular}
\end{table}

The absolute values of the $g$~factors in the Ru isotopes as presented in Table \ref{tab:Ruresults} were
determined by use of the Rutgers parametrization \cite{shu80}, which was demonstrated in the previous section (\ref{sect:convCd}) to be applicable for ions traversing iron hosts under the conditions of this measurement. The uncertainties shown in Table \ref{tab:Ruresults} correspond to the uncertainties in the relative $g$~factors. Uncertainties on the absolute values of the $g$~factors will be discussed in sect. \ref{sect:adopted} below, where these measurements will be combined with the measurements in inverse kinematics.

\subsection{Inverse kinematics: Ru, Pd and Cd isotopes} \label{sect:inv}

\begin{table}[h]
\caption{Summary of measurements in inverse kinematics. $E_B$ and $I_B$ are the beam energy and intensity.}
\label{tab:experiments}
\begin{ruledtabular}
\begin{tabular}{lccll}
 Beam & $E_B$ & $I_B$ & Measurement\footnotemark[1]; Purpose  &
 Duration\\
 & (MeV) & (enA) &  & (h) \\
 \hline
%
%
\multicolumn{5}{c}{Target I}\\
$^{102}$Ru$^{16+}$ & 245 & 5 & $\epsilon$; $g$($^{102}$Ru)/$g$($^{108}$Pd) & 2.5  \\

$^{108}$Pd$^{15+}$ & 230 & 5 & $\epsilon$; $B_{\rm TF}(v)$ & 2  \\
$^{108}$Pd$^{15+}$ & 245 & 5 & $\epsilon$; $B_{\rm TF}(v)$ & 2.25  \\
$^{108}$Pd$^{17+}$ & 260 & 5 & $\epsilon$; $B_{\rm TF}(v)$ & 1.75  \\
$^{114}$Cd$^{16+}$ & 240 & 5 & $\epsilon$; $g$($^{114}$Cd)/$g$($^{108}$Pd) & 5.5  \\
\multicolumn{5}{c}{Target II}\\
$^{96}$Ru$^{16+}$ & 240 & 1.5 & $\epsilon$; $g$($^{96}$Ru) & 10  \\
$^{98}$Ru$^{16+}$ & 240 & 0.5 & $\epsilon$; $g$($^{98}$Ru) & 25  \\
$^{100}$Ru$^{16+}$ & 240 & 4 & $\epsilon$; $g$($^{100}$Ru) & 4.5  \\
$^{102}$Ru$^{16+}$ & 240 & 3 & $\epsilon$; $g$($^{102}$Ru) & 1  \\
$^{104}$Ru$^{16+}$ & 240 & 5 & $\epsilon$; $g$($^{104}$Ru) & 2  \\
$^{102}$Pd$^{16+}$ & 245 & 0.8 & $\epsilon$; $g$($^{102}$Pd) & 8  \\
$^{104}$Pd$^{16+}$ & 245 & 6 & $\epsilon$; $g$($^{104}$Pd) & 2.2  \\
$^{106}$Pd$^{16+}$ & 245 & 8 & $\epsilon$; TF calibration & 3.3  \\
$^{108}$Pd$^{16+}$ & 245 & 6 & $\epsilon$,$W(\Theta)$ ; $g$($^{108}$Pd)  & 3.5  \\
$^{110}$Pd$^{16+}$ & 245 & 3 & $\epsilon$; $g$($^{110}$Pd) & 1.7  \\
$^{106}$Cd$^{16+}$ & 240 & 2  & $\epsilon$; $g$($^{106}$Cd)   & 9.6  \\
$^{108}$Cd$^{16+}$ & 240 & 2  & $\epsilon$; $g$($^{108}$Cd)   & 11  \\
$^{112}$Cd$^{16+}$ & 240 & 1.4& $\epsilon$; $g$($^{106,108}$Cd)/$g$($^{112}$Cd) & 18  \\
\multicolumn{5}{c}{Target III}\\
$^{98}$Mo$^{16+}$ & 240 & 2.2 & $\epsilon$; TF calibration & 18  \\
$^{96}$Ru$^{16+}$ & 240 & 2.5 & $\epsilon$; $g$($^{96}$Ru) & 18  \\
$^{100}$Ru$^{16+}$ & 240 & 2.5 & $\epsilon$; $g$($^{100}$Ru) & 12  \\
$^{102}$Ru$^{16+}$ & 240 & 2.5 & $\epsilon$; $g$($^{102}$Ru) & 15  \\
$^{104}$Ru$^{16+}$ & 240 & 2.5 & $\epsilon$; $g$($^{104}$Ru) & 10  \\
$^{106}$Pd$^{16+}$ & 240 & 2 & $\epsilon$; TF calibration  & 12  \\
\end{tabular}
\footnotetext[1]{~$\epsilon$: transient-field precession;
$W(\Theta)$ : angular correlation.}
\end{ruledtabular}

\end{table}

As summarized in Table~\ref{tab:experiments}, transient-field
measurements in inverse kinematics were performed on all of the
stable even Ru and Pd isotopes, and $^{106,108,112,114}$Cd, using
$\sim 2.3$~MeV/$A$ beams from the 14UD Pelletron. The beam
intensities ranged from $\sim 0.5$~pnA for $^{106}$Pd to $\sim
0.03$~pnA for $^{98}$Ru. Negative ion beams of the Ru and Pd
isotopes were produced from natural metal powder pressed into a
standard copper cathode. CdO$^-$ beams were produced from cadmium
oxide - natural for the $^{112}$Cd and $^{114}$Cd beams, and partially enriched for
the $^{106}$Cd and $^{108}$Cd beams. Beams of $^{98}$MoO$_2^-$ ions were obtained from a metallic Mo cathode in the presence of O$_2$ gas.

For these inverse kinematics experiments the ANU
Hyperfine Spectrometer was configured with a forward array of three
particle detectors; the apparatus and experimental procedures were
similar to those in our recent work on the Fe isotopes
\cite{eas09,eas09a}.

The first two targets (labeled I and II) used for these measurements consisted
of C layers on 6.1~mg/cm$^2$ thick copper-backed gadolinium foils.
After rolling and annealing under vacuum, a thin 0.04~mg/cm$^2$
layer of copper was evaporated onto the beam-facing side (front) of
the gadolinium foil to assist the adhesion of the C layer, and a
thicker 5.5~mg/cm$^2$ layer of copper was evaporated on the back.
The layer of carbon, 0.4~mg/cm$^2$ thick, was added to the front of
the target by applying a suspension of carbon powder in isopropyl
alcohol. Additional copper foil (4.5~mg/cm$^2$) was placed behind
the target to stop the beam. The target was cooled below 5 K, both
to minimize the effect of beam heating, and to maximize the
magnetization of the gadolinium layer of the target.
Although they were nominally the same, Target I gave
somewhat larger precessions for $^{108}$Pd ions at 245 MeV than
Target II. We attribute this difference to variations in the
effective thickness of the target layers at the beam spot. It is also possible that the magnetization differs, despite the targets having been prepared from the same annealed gadolinium foil. An external magnetic field of 0.09~T was applied perpendicular to the $\gamma$-ray detection plane to magnetize the gadolinium layer of the target. This field was reversed reversed periodically throughout the measurements.

Additional experiments were performed with a third target (labeled Target III) which consisted of enriched $^{26}$Mg, 0.45 mg/cm$^2$ thick, evaporated onto gadolinium, 3.2 mg/cm$^2$ thick, which was followed by layers of nickel (0.01 mg/cm$^2$) and copper (5.4 mg/cm$^2$). This target had previously been used for similar measurements at Yale by Taylor {\em et al.} \cite{Taylor2010}.

Level properties of the beam ions and the reaction kinematics for target ions ($^{12}$C or $^{26}$Mg) scattered into the outer particle detectors (average scattering angle 20.7$^\circ$) are summarized in Table~\ref{tab:inverse-kinem}.

\begin{table*}
\caption{Level properties and reaction kinematics for Mo, Ru, Pd and Cd beam ions recoiling in gadolinium after Coulomb excitation on target ions in inverse kinematics during the $g$-factor measurements. The reaction kinematics are shown for recoiling $^{12}$C (Target I and II) or $^{26}$Mg (Target III) ions detected in the outer
particle counters (average scattering angle 20.7$^\circ$). See Table
\protect \ref{tab:Cdkinem} and text.} \label{tab:inverse-kinem}
\begin{ruledtabular}
\begin{tabular}{cccccccccc}
Isotope & $E_B$ & $E(2^+_1)$ & $\tau(2^+_1)$ & $\langle E_i \rangle$
& $\langle E_e \rangle$ & $\langle v_i/v_0 \rangle$ & $\langle
v_e/v_0
\rangle$ & $\langle v/v_0 \rangle$ & $t_{\rm Gd}$ \\
 & (MeV) & (keV) & (ps) & (MeV) & (MeV) & & & & (fs) \\ \hline
\multicolumn{10}{c}{ Targets I and II} \\
$^{96}$Ru  &240& 832.6 &  4.1 & 136.5 & 29.7 & 7.57 & 3.53 & 5.38 & 576    \\
$^{98}$Ru  && 652.4 &  8.8 & 137.6 & 30.9 & 7.52 & 3.56 & 5.34 & 614    \\
$^{100}$Ru && 539.6 & 18.2 & 138.8 & 32.0 & 7.48 & 3.60 & 5.32 & 638    \\
$^{102}$Ru && 475.1 & 26.6 & 139.9 & 33.3 & 7.44 & 3.63 & 5.32 & 644    \\
$^{104}$Ru && 358.0 & 83.4 & 141.0 & 33.5 & 7.39 & 3.65 & 5.31 & 653    \\
$^{102}$Pd &245& 556.4 & 16.6 & 142.4 & 32.3 & 7.49 & 3.57 & 5.32 & 637    \\
$^{104}$Pd && 555.8 & 14.4 & 143.6 & 33.6 & 7.44 & 3.61 & 5.33 & 633    \\
$^{106}$Pd && 511.9 & 17.6 & 144.7 & 34.8 & 7.42 & 3.64 & 5.33 & 633    \\
$^{108}$Pd && 434.0 & 34.7 & 145.8 & 35.9 & 7.38 & 3.66 & 5.32 & 647    \\
$^{110}$Pd && 373.8 & 63.5 & 146.9 & 37.1 & 7.34 & 3.69 & 5.31 & 652    \\
$^{106}$Cd &240& 632.6 & 9.8 & 140.6 & 31.4 & 7.31 & 3.46 & 5.16 & 643    \\
$^{108}$Cd && 633.0 &  9.3 & 141.7 & 32.6 & 7.27 & 3.49 & 5.17 & 641    \\
$^{112}$Cd && 617.5 &  8.9 & 143.9 & 34.9 & 7.19 & 3.54 & 5.18 & 638    \\
$^{114}$Cd && 558.5 & 13.7 & 144.8 & 35.9 & 7.16 & 3.56 & 5.16 & 652    \\
\multicolumn{10}{c}{ Target III} \\
$^{98}$Mo &240&787.4 & 5.1 & 89.3 & 37.8 & 6.06 & 3.94 & 4.93 & 367    \\
$^{96}$Ru  && 832.6 &  4.1 & 87.1 & 34.9 & 6.05 & 3.83 & 4.86 & 368    \\
$^{100}$Ru && 539.6 & 18.2 & 90.1 & 37.7 & 6.03 & 3.90 & 4.88 & 385    \\
$^{102}$Ru && 475.1 & 26.6 & 91.6 & 39.1 & 6.02 & 3.93 & 4.89 & 385 \\
$^{104}$Ru && 358.0 & 83.4 & 93.1 & 40.1 & 6.01 & 3.96 & 4.90 & 386
\\
$^{106}$Pd && 511.9 & 17.6 & 94.1 & 40.3 & 5.98 & 3.92 & 4.87 & 384    \\
\end{tabular}
\end{ruledtabular}
\end{table*}

\begin{figure*}[ht]
  \includegraphics[width=0.90\textwidth]{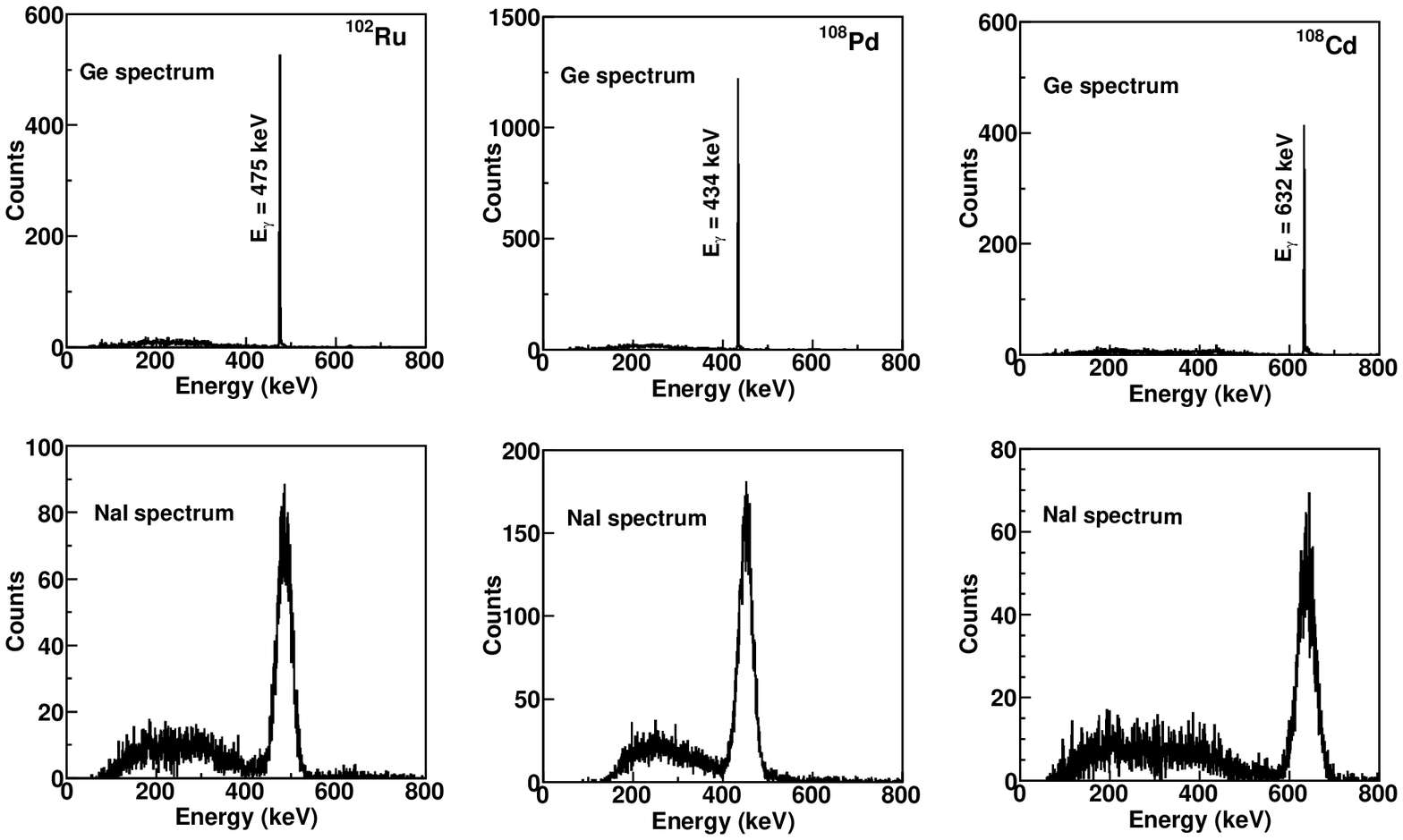}
  \caption{Examples of $\gamma$-ray spectra observed at
  $+65^\circ$ and $+115^\circ$ to the
beam axis in coincidence with C ions forward scattered into the
outer particle counters. These spectra represent all of the data for
the field up direction, obtained during the precession measurements
on $^{102}$Ru, $^{108}$Pd and $^{108}$Cd using Target II.}
  \label{fig:inverse-specta}
\end{figure*}

The de-exciting $\gamma$ rays from the Mo, Ru, Pd and Cd isotopes were
measured in coincidence with forward-scattered target ions detected
by an array of three silicon photodiode detectors downstream from
the target, arranged in a vertical stack as described in
\cite{eas09}. For the measurements in inverse kinematics, two 50\%
efficient HPGe detectors and two 12.7~cm by 12.7~cm NaI detectors
were placed in pairs at $\pm 65^\circ$ and $\pm 115^\circ$ to the
beam axis, respectively. The target-detector distances were again
set such that the detector crystals all subtended a half angle of
$18^\circ$. Figure \ref{fig:inverse-specta} shows examples of
coincidence $\gamma$-ray spectra observed in the Ge detector at
$+65^\circ$ to the beam direction and the NaI detector at
$+115^\circ$ to the beam.

\begin{figure}[ht]
  \includegraphics[width=0.45\textwidth]{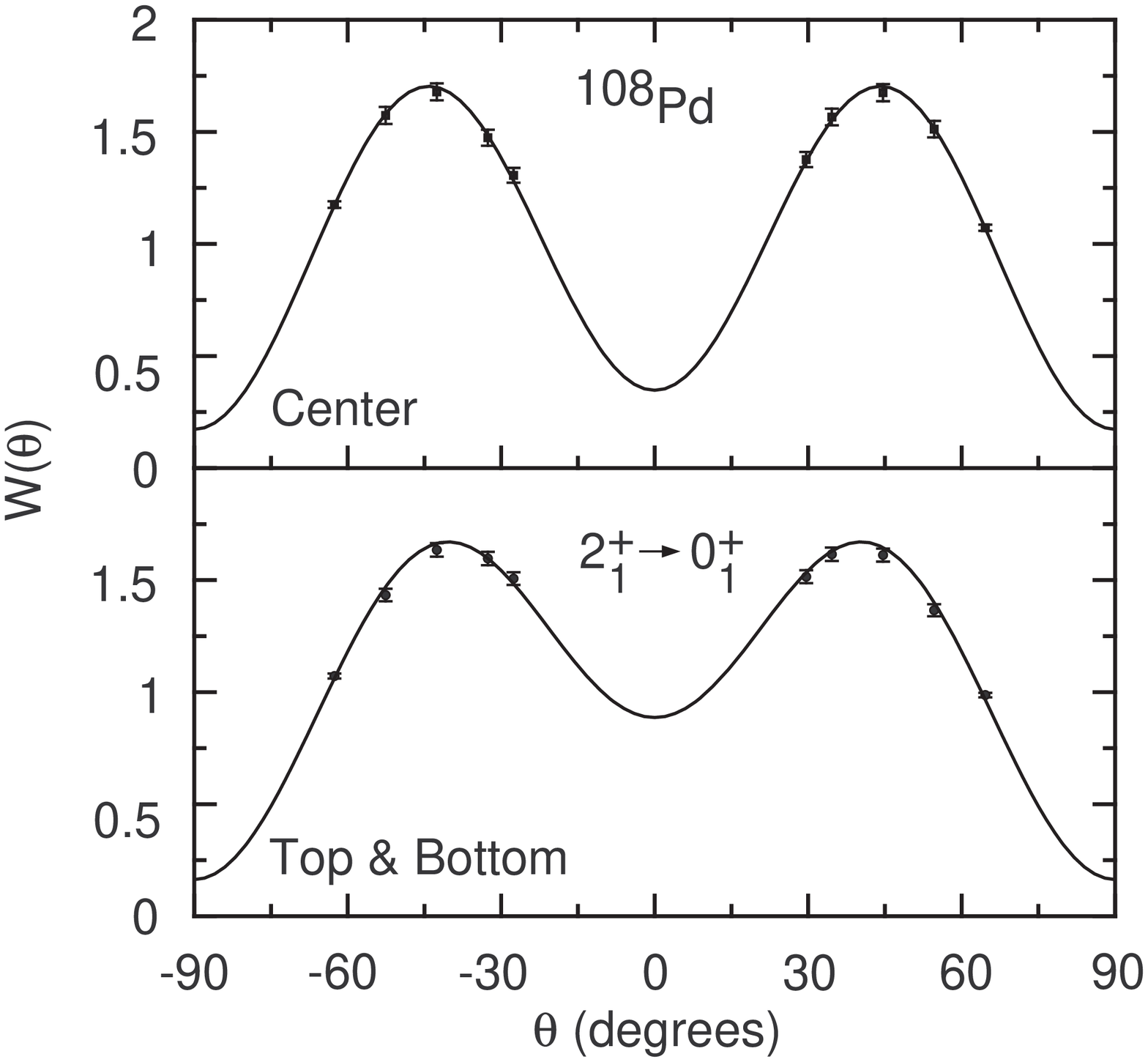}
\caption{Measured and calculated angular correlations for $^{108}$Pd
in inverse kinematics. The upper panel is for the center particle
detector (average $^{12}$C scattering angle 0$^\circ$); the lower
panel is for the top and bottom (outer) particle detectors (average
$^{12}$C scattering angle 20.7$^\circ$).}
  \label{fig:Pd108AD}
\end{figure}

Angular correlations were measured for $^{108}$Pd beams at 245 MeV,
in a sequence of runs of about 20 min duration. The backward-placed
NaI detectors were kept at $\pm 115^\circ$, to normalize the number
of counts, while the angular correlation was sampled with the two
forward Ge detectors at $\pm 30.5^\circ$, $\pm 35^\circ$, $\pm
45^\circ$, $\pm 55^\circ$ and $\pm 65^\circ$, in turn. The results
of these measurements are compared with the calculated angular
correlations in Fig.~\ref{fig:Pd108AD}. Calculated angular
correlations were used for the analysis of the precession data. At
the relatively low beam energy of $\sim 2.3$ MeV/nucleon multiple
excitation is very small and the angular correlations for all
isotopes are effectively identical \cite{eas09a}. The slope
parameters are $S(65^\circ)=- 2.69$ rad$^{-1}$ for the outer particle counters, and $S(65^\circ)= - 2.60$ rad$^{-1}$ for the center detector.

\begin{figure}[ht]
  \includegraphics[width=0.45\textwidth]{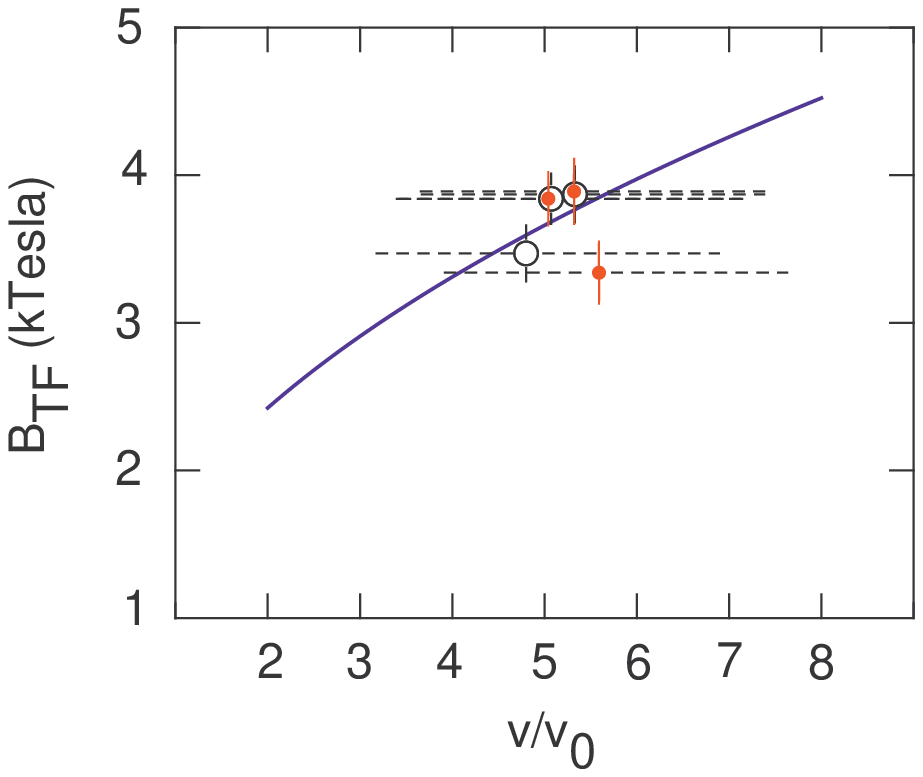}
\caption{(Color online) Measured transient-field strength for $^{108}$Pd ions
traversing the gadolinium layer of Target I. Dotted horizontal lines
indicate the velocity range over which the transient field is
sampled. The average transient-field strength is plotted at the
average ion velocity. Open symbols correspond to the detection of $^{12}$C ions in the central detector while filled symbols indicate detection in the outer detectors with an average scattering angle of 20.7$^\circ$. The solid line shows the $v^{0.45}$ velocity
dependence of the Rutgers parametrization \protect \cite{shu80} scaled to best fit these data.}
  \label{fig:Pd108TFstrength}
\end{figure}

The velocity dependence of Pd ions traversing gadolinium was
investigated through measurements on 230, 245 and 260 MeV $^{108}$Pd
ions traversing Target I. The results are summarized in
Table~\ref{tab:TFveldep}. In these measurements the Pd ions sample
the transient field over a velocity range that extends beyond that
covered in the $g$-factor measurements. Between the three runs at
different beam energies there is an overlap of the velocity range
sampled by the $^{108}$Pd ions in consecutive measurements due to
the difference in scattering conditions for the center ($\langle
\theta_{\rm C} \rangle = 0^\circ$) and outer ($\langle \theta_{\rm
O} \rangle = 20.7^\circ $) particle detectors. Agreement between the
observed precessions for the measurements that span similar velocity
ranges is excellent. Fig.~\ref{fig:Pd108TFstrength} compares the
velocity dependence of the average transient field strength for Pd ions traversing gadolinium with
that of the Rutgers parametrization. While the field strength parameter $a_{\rm TF}$ has to be scaled up by a factor of about 1.4 for Target I, the data are nevertheless consistent with a $v^{0.45}$ dependence over the range applicable for the present $g$-factor measurements. To evaluate relative $g$~factors, we have
therefore used the Rutgers parametrization to determine the
scaling ratios described in sect. \ref{sect:experiment}, which are needed to correct for differences in the velocity
range over which the different isotopes experience the transient
field.

\begin{table*}
\caption{Reaction kinematics, precessions and average
transient-field strengths for $^{108}$Pd. See Tables \ref{tab:Cdkinem} and \ref{tab:Cdresults} and text. $\langle \theta_{\rm carbon} \rangle$ is the average scattering angle of the carbon target ions.} \label{tab:TFveldep}
\begin{ruledtabular}
\begin{tabular}{cccccccccc}
$E_B$ & $\langle \theta_{\rm carbon} \rangle$ & $\langle E_i
\rangle$ & $\langle E_e \rangle$ & $\langle v_i/v_0 \rangle$ &
$\langle v_e/v_0 \rangle$ & $\langle v/v_0 \rangle$ & $t_{\rm Gd}$ &
$\Delta \Theta _{\rm obs}$ &
$\langle B_{\rm TF} \rangle$ \\

(MeV)& $(^\circ)$ & (MeV) & (MeV) &  &  & & (fs) & (mrad) & (ktesla) \\
\hline
230 & 0 & 127.1 & 27.1 & 6.89 & 3.18 & 4.80 & 713 & $-40.99(2.23)$ &
3.47(19)\\
230 & 20.7& 135.7 & 31.0 & 7.12 & 3.40 & 5.04 & 682 & $-43.40(2.01)$
& 3.84(18)\\
245 & 0 & 136.6 & 31.6 & 7.14 & 3.43 & 5.07 & 676 & $-42.99(1.86)$ &
3.84(17)\\
245 & 20.7& 145.8 & 35.9 & 7.38 & 3.66 & 5.32 & 647 & $-41.66(2.34)$
& 3.89(22)\\
260 & 0 & 146.1 & 36.2 & 7.38 & 3.67 & 5.33 & 644 & $-41.23(2.01)$ &
3.87(19)\\
260 & 20.7 & 156.0 & 41.2 & 7.63 & 3.92 & 5.59 & 616 &
$-34.06(2.18)$ & 3.34(21)\\
\end{tabular}
\end{ruledtabular}
\end{table*}

The absolute scale of the experimental $g$ factors has been set here
by reference to previous measurements on $^{106}$Pd by the
external field and radioactivity techniques \cite{joh72,joh68}. We have followed the recommendation of
Johansson {\it et al.} \cite{joh72} and disregarded the measurements
that used iron hosts, which gave evidence of having slightly reduced
hyperfine fields. The adopted $g$ factor is then the weighted
average of their measurement using a cobalt host \cite{joh72} and an earlier
external field measurement with which it agrees \cite{joh68}.
After making a small correction for a more recent level lifetime
($\tau = 17.6(9)$~ps \cite{Raman}), we obtain $g = +0.393(23)$, with
the uncertainty dominated by the uncertainty in the lifetime. The
most recent Nuclear Data evaluation \cite{ndsPd106} for $^{106}$Pd
includes iron-host data \cite{bow68,sin70}. If these data are
included in the average (after being adjusted to correspond to the
same adopted lifetime), the resultant value, $g = +0.378(21)$,
is smaller than, but remains consistent with, our adopted value.

\begin{figure}[ht]
  \includegraphics[width=0.45\textwidth]{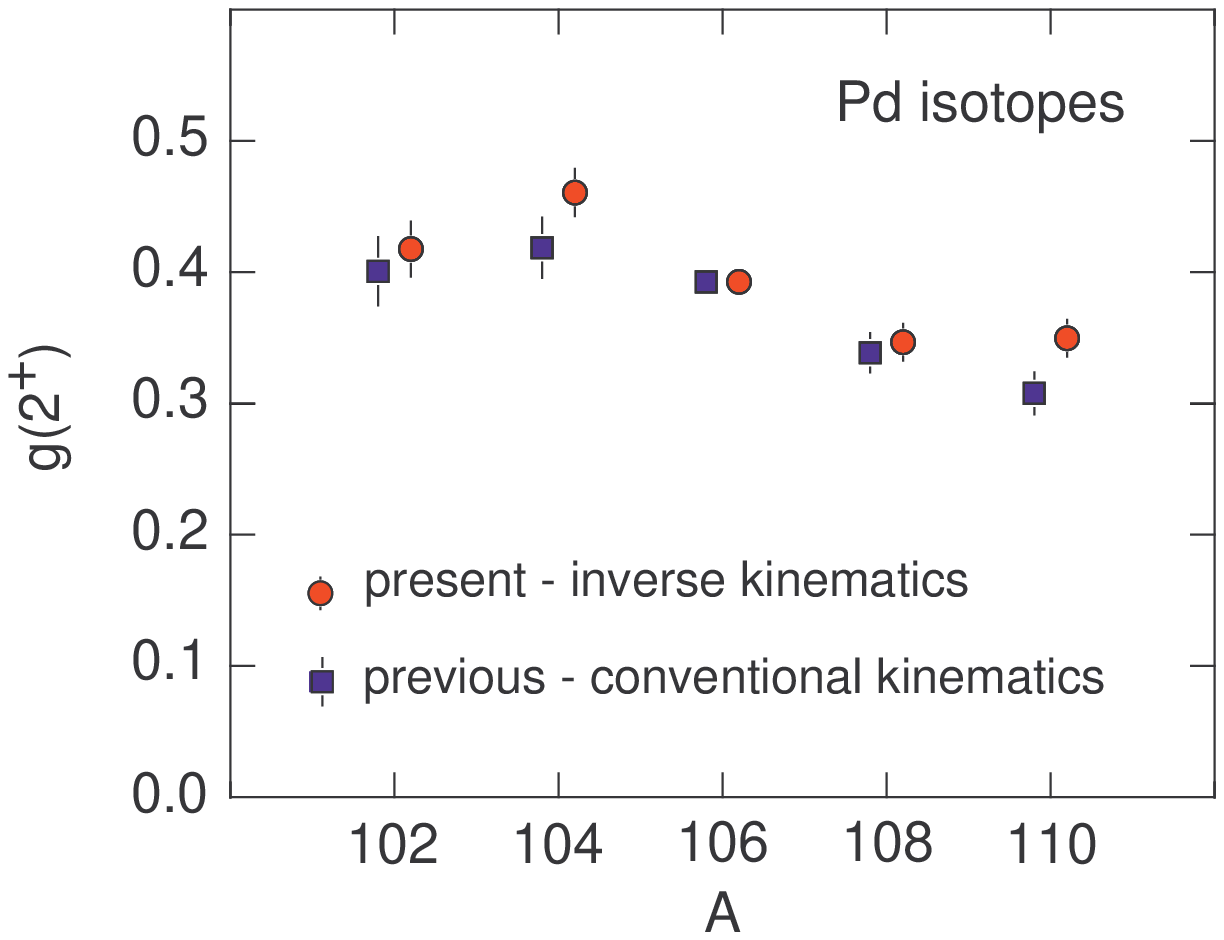}
\caption{Comparison of transient-field measurements on the Pd isotopes. The present measurements in inverse kinematics are compared with previous measurements in conventional kinematics \protect
\cite{bre80,tho85,stu84,lam89}. The data are shown relative to the present adopted $g(2^+)$ value in $^{106}$Pd. }
  \label{fig:pdcomparison}
\end{figure}

Precession and $g$~factor results are summarized in
Table~\ref{tab:inv-results}. Figure~\ref{fig:pdcomparison} shows a
comparison of the present and previous data for relative $g$~factors in the Pd isotopes. The previous measurements used the transient-field technique in conventional kinematics whereas the present measurements use inverse kinematics. The consistency of the data in Fig.~\ref{fig:pdcomparison} is important because transient-field measurements in inverse kinematics, like those reported here, are reaching a high level of statistical precision, to the point where uncontrolled systematic effects associated with changing the beam species might become evident.

\begin{table}[t]
\caption{Results of $g$-factor measurements in inverse kinematics. $\langle \Delta \Theta \rangle_{\rm obs}$ is the average observed experimental precession angle. The $g$~factors are referenced to $g(2^+_1;^{106}{\rm Pd} =+0.393$ and assigned uncertainties that reflect the uncertainties in the measured precession angles. Adopted $g$~factors, which include uncertainties in the reference $g$~factor, are presented in sect.~\ref{sect:adopted}.} \label{tab:inv-results}
\begin{ruledtabular}
\begin{tabular}{ccc}
Nuclide & $\langle \Delta \Theta\rangle _{\rm obs}$ &
$g$~\footnotemark[1] \\
 & (mrad) &  \\
\hline
\multicolumn{3}{c}{Target I}\\
$^{102}$Ru  &   $-50.68(148)$  & $+0.436(13)$ \\
$^{108}$Pd  &   $-42.48(146)$  & $+0.347(12)$ \\
$^{114}$Cd  &   $-41.43(175)$  & $+0.313(14)$ \\

\multicolumn{3}{c}{Target II}\\
$^{96}$Ru   &   $-40.94(223)$  & $+0.440(24)$ \\
$^{98}$Ru   &   $-40.18(217)$  & $+0.414(22)$ \\
$^{100}$Ru  &   $-42.95(136)$  & $+0.431(14)$ \\
$^{102}$Ru  &   $-45.01(191)$  & $+0.449(19)$ \\
$^{104}$Ru  &   $-44.20(138)$  & $+0.436(14)$ \\
 \\
$^{102}$Pd  &   $-43.84(196)$  & $+0.418(19)$ \\
$^{104}$Pd  &   $-48.12(140)$  & $+0.461(14)$ \\
$^{106}$Pd  &   $-41.19(92)$   & $+0.393(9)$ \\
$^{108}$Pd  &   $-36.79(126)$  & $+0.347(12)$ \\
$^{110}$Pd  &   $-37.23(119)$  & $+0.350(11)$ \\
 \\
$^{106}$Cd  &   $-43.08(228)$  & $+0.393(21)$ \\
$^{108}$Cd  &   $-42.02(220)$  & $+0.389(20)$ \\
$^{112}$Cd  &   $-39.80(232)$  & $+0.365(21)$ \\
\multicolumn{3}{c}{Target III}\\		
$^{98}$Mo   &   $-30.42(223)$  & $+0.459(34)$ \\
$^{96}$Ru   &   $-29.70(169)$  & $+0.432(24)$ \\
$^{100}$Ru  &   $-29.89(141)$  & $+0.411(20)$ \\
$^{102}$Ru  &   $-31.21(97)$   & $+0.428(14)$ \\		
$^{104}$Ru  &   $-26.99(150)$  & $+0.369(14)$ \\				
$^{106}$Pd  &   $-30.03(148)$  & $+0.393(20)$ \\
\end{tabular}
\end{ruledtabular}
\end{table}

\subsection{Adopted $g$ factors} \label{sect:adopted}

This section gives a summary of the present and previous $g(2^+_1)$ values in Mo, Ru, Pd and Cd
isotopes.

Table \ref{tab:Mo-data} summarizes the previous data on the Mo isotopes from Refs. \cite{Man2001,smith05}, along with the values adopted for the following comparison with theory. The new measurement for $g(2^+_1)$ in $^{98}$Mo, relative to $^{106}$Pd, is included in the summary of adopted $g$ factors.

Table \ref{tab:Ru-data} summarizes the $g$~factors in the Ru isotopes. Further explanation is required concerning the method used to combine the results of the three independent sets of measurements on the Ru isotopes (one in conventional kinematics and two in inverse kinematics). On one hand, the two measurements in inverse kinematics with gadolinium as the ferromagnetic host were calibrated relative to measurements on $^{106}$Pd (and $^{98}$Mo) performed with the same target. On the other hand, the measurement in conventional kinematics, with an iron host, was calibrated (sect. \ref{sect:convRu}) using the Rutgers parametrization \cite{shu80}. In this situation it is difficult to properly combine the data sets and propagate the errors by the usual procedure of working with one or two $g$-factor ratios. Nevertheless, the mathematical relationships relating the data sets are simple.

The procedure adopted to combine these measurements began by writing down the relationships between all of the experimental `knowns', namely the measured precession angles and previously determined $g$~factor values, and the experimental `unknowns', i.e. the transient-field strengths and the $g$~factors to be extracted from the data. To illustrate the procedure, it is most convenient to work with corrected experimental precession angles, $\Delta \Theta(A)$, as defined in Eq.~(\ref{eq:dth-corr}) and the integral transient-field strengths, $\phi(\infty)$, as defined in Eq.~(\ref{eq:phi-infty}). The 6 precession measurements on Target II are then related by $\Delta \Theta_{II}(A)=g(A) \phi_{II}$, where $A$ denotes the 6 even-even nuclei $^{96-104}$Ru and $^{106}$Pd; $\phi_{II}$ is the same for all measurements on Target II. Similarly, the 6 precession measurements on Target III are related by $\Delta \Theta_{III}(A)=g(A) \phi_{III}$, where $A$ now denotes $^{96,100-104}$Ru, $^{98}$Mo and $^{106}$Pd. The 3 experimental precessions measured in conventional kinematics are related by $\Delta \Theta_{\rm conv}(A)=g(A) \phi_{\rm conv}$, where $A$ denotes $^{100-104}$Ru. With these data alone, only ratios of $g$~factors can be obtained. Additional data must be used to obtain the absolute values of the $g$~factors or, alternatively, the values of $\phi_{II}$, $\phi_{III}$ and $\phi_{\rm conv}$. As noted above, for this purpose we have used the previously measured $g$~factors in both $^{106}$Pd and $^{98}$Mo, along with the Rutgers parametrization \cite{shu80} to determine $\phi_{\rm conv}$. The parametrization of the field strength, and hence $\phi_{\rm conv}$, was assigned a 10\% uncertainty (i.e. $a_{\rm TF}=16.9\pm1.7$~T in Eq.~(\ref{eq:Bparam})).

The set of equations relating the experimental precessions, the $g$~factors, the integrated transient-field strengths, and the previous data to be used for field calibration, were then used as the basis for a chi-squared fit to determine the Ru $g$~factors. The fitting procedure gives the correct average values for the Ru $g$~factors and their associated experimental uncertainties, including the uncertainty in the transient-field calibration.

 The data included in the fit were: the precession data for all three measurements, the previous experimental $g$ factors in $^{106}$Pd and $^{98}$Mo, and the scale parameter of the transient field for the Rutgers parametrization applied to iron hosts.  There were therefore 18 data values in the fit (15 $\Delta \Theta$ values, two independently measured $g$~factors and one transient-field scale factor). Ten parameters were extracted from the fit: the five $g$~factors for the even Ru isotopes between $^{96}$Ru and $^{104}$Ru, the transient-field strengths in the three measurements, and the $g$~factors of $^{106}$Pd and $^{98}$Mo. Note that new values of the latter two $g$~factors were output fit parameters while their previous experimental values were input data for the fit.

\begin{figure}[ht]
  \includegraphics[width=0.45\textwidth]{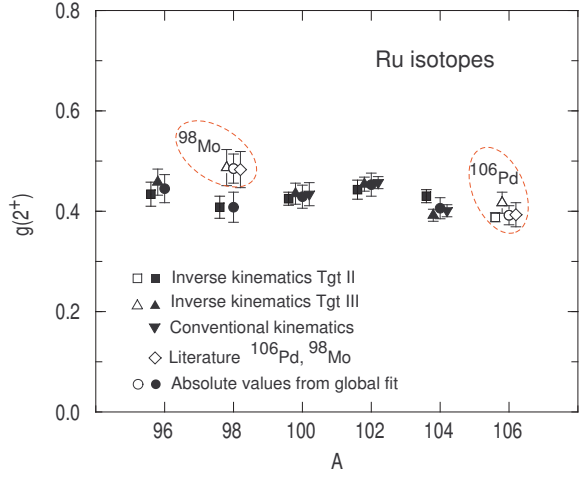}
\caption{(Color online) Comparison of $g$ factors from the three measurements on the Ru isotopes, calibration values from the literature, and the average values obtained from the global fit. See text.}
  \label{fig:gfit}
\end{figure}

 The chi-squared per degree of freedom was 0.95. The results of this fit procedure and the consistency of the three data sets are shown in Fig.~\ref{fig:gfit}. The field calibration, and hence the absolute values of the $g$~factors, was determined predominantly by the previous $g$~factor in $^{106}$Pd. Indeed, the value of $g(2^+)$ in $^{106}$Pd returned by the fit did not differ significantly in either magnitude or precision from the adopted previous value. For $^{98}$Mo, however, an improved $g(2^+)$ value was obtained, and is reported in Table~\ref{tab:Mo-data}. It is worth noting that the transient field strength for Ru in iron hosts agreed with the Rutgers parametrization \cite{shu80} to within 5\%, whereas for the measurements on gadolinium hosts the transient field strength was of order 30\% stronger than the prediction of the Rutgers parametrization.

As can be seen from Table \ref{tab:Ru-data} the present $g$~factor for $^{102}$Ru is 20\% higher than that obtained by Johansson {\em et al.} from perturbed angular correlation measurements on a radioactive source alloyed with an iron host \cite{joh72}. They assumed a static hyperfine field of $B_{\rm hf}=50.3 \pm 0.9$~T. Their paper discusses at length, however, the possible processes that introduce systematic errors in this type of measurement, which usually cause a reduction in the effective hyperfine field. The present $g$-factor measurement implies that in Ref.~\cite{joh72} the effective static field for Ru impurities in iron was $B_{\rm hf}=-40\pm3$~T. Our results are in better agreement with the radioactivity measurements of Auerbach {\em et al.} \cite{Auerbach66} who took $B_{\rm hf}=-44\pm3$~T.

Smith {\em et al.} \cite{smith04,smith05} adopted a hyperfine field for Ru in iron of $B_{\rm hf}=-47.8\pm0.1$~T. The above discussion suggests that this may be an over-estimate. However, aside from noting that there may be grounds to increase the previously reported $g$~factors of the neutron-rich isotopes  \cite{smith04,smith05} by about 20\%, these values are reported without adjustment in Table \ref{tab:Ru-data}.

Finally, it is to be noted that Taylor {\em et al.} \cite{Taylor2010} performed transient-field $g$-factor measurements on the Ru isotopes in parallel with the present work. Their adopted $g$~factors agree within the experimental uncertainties with those reported here.

\begin{table}
\caption{$g(2^+_1)$ values in the Mo isotopes.}
\label{tab:Mo-data}
\begin{ruledtabular}
\begin{tabular}{cccc}
Nuclide & \multicolumn{3}{c}{$g(2^+)$} \\ \cline{2-4}
 & Previous & Present & Adopted \\
\hline      \\
 $^{92}$Mo  & +1.15(14)  \footnotemark[1] & -        & +1.15(14) \\
 $^{94}$Mo  & +0.308(43) \footnotemark[1] & -        & +0.308(43)\\
 $^{96}$Mo  & +0.394(31) \footnotemark[1] & -        & +0.394(31)\\
 $^{98}$Mo  & +0.483(36) \footnotemark[1] &+0.485(29)& +0.485(29)\\
$^{100}$Mo  & +0.471(33) \footnotemark[1] & -        & +0.471(33)\\
$^{102}$Mo  & +0.42(7)   \footnotemark[2] & -        & +0.42(7)\\
$^{104}$Mo  & +0.27(2)   \footnotemark[3] & -        & +0.27(2)\\
$^{106}$Mo  & +0.21(2)   \footnotemark[3] & -        & +0.21(2)\\
$^{108}$Mo  & +0.5(3)    \footnotemark[3] & -        & +0.5(3)\\
                & \\
\end{tabular}
\footnotetext[1]{Adopted value in \protect \cite{Man2001}.}
\footnotetext[2]{\protect \cite{men85}.}
\footnotetext[3]{\protect \cite{smith05}.}
\end{ruledtabular}
\end{table}

\begin{table}
\caption{$g(2^+_1)$ values in the Ru isotopes. } \label{tab:Ru-data}
\begin{ruledtabular}
\begin{tabular}{cccc}
Nuclide & \multicolumn{3}{c}{$g(2^+)$} \\ \cline{2-4}
 & Previous & Present & Adopted \\
\hline                & \\
 $^{96}$Ru  &                          &  +0.445(28) & +0.445(28)\\
 $^{98}$Ru  &+0.4(3)  \footnotemark[1] &  +0.408(32) & +0.408(32)\\
$^{100}$Ru  &+0.46(5) \footnotemark[2] &  +0.429(23) & +0.429(23)\\
$^{102}$Ru  &+0.37(8) \footnotemark[2]\\
            &+0.354(21) \footnotemark[3]& +0.453(23) & +0.453(23)\\
$^{104}$Ru  &+0.41(5) \footnotemark[1] &  +0.406(21) & +0.406(21)\\
$^{106}$Ru  &+0.28(13)\footnotemark[4] &             & +0.28(13) \footnotemark[5] \\
$^{108}$Ru  &+0.28(4) \footnotemark[4] &             & +0.28(4)  \footnotemark[5] \\
$^{110}$Ru  &+0.42(6) \footnotemark[4] &             & +0.42(6)  \footnotemark[5] \\
$^{112}$Ru  &+0.44(9) \footnotemark[4] &             & +0.44(9) \footnotemark[5] \\
\end{tabular}
\end{ruledtabular}
 \footnotetext[1]{From \protect \cite{HUB74}. }
 \footnotetext[2]{From \protect \cite{Auerbach66}, updated for lifetime in \protect \cite{Raman}.}
 \footnotetext[3]{From \protect \cite{joh72}, updated for lifetime in \protect \cite{Raman}.}
 \footnotetext[4]{From \protect \cite{smith05}.}
 \footnotetext[5]{As discussed in the text, there may be evidence that these values should be increased by a factor of 1.2.}
\end{table}

\begin{table}
\caption{$g(2^+_1)$ values in the Pd isotopes. Uncertainties in square brackets are relative errors for the
sequence of isotopes; errors in round brackets include the uncertainty in the transient-field calibration.} \label{tab:Pd-data}
\begin{ruledtabular}
\begin{tabular}{cccc}
Nuclide & \multicolumn{3}{c}{$g(2^+)$} \\ \cline{2-4}
 & \multicolumn{1}{c}{\rm previous} & \multicolumn{1}{c}{\rm present} & \multicolumn{1}{c}{\rm adopted} \\
\hline                & \\
$^{102}$Pd  & +0.401 [26]   \footnotemark[1] & +0.418[21] & 0.411(30) \\
$^{104}$Pd  & +0.419 [0.23] \footnotemark[1] & +0.461[18] & 0.446(30) \\
$^{106}$Pd  & +0.393 [0]    \footnotemark[1] & +0.393[0]  & 0.393(23) \\
$^{108}$Pd  & +0.339 [15]   \footnotemark[1] & +0.347[14] & 0.343(22) \\
$^{110}$Pd  & +0.308 [16]   \footnotemark[1] & +0.350[14] & 0.333(21) \\
$^{112}$Pd                           & \\
$^{114}$Pd  & +0.24(13)   \footnotemark[2] &              & +0.24(13)   \\
$^{116}$Pd  & +0.2(1)     \footnotemark[2] &              & +0.2(1)   \\
                & \\
\end{tabular}
\end{ruledtabular}
 \footnotetext[1]{Average of relative $g$-factor measurements in \protect \cite{bre80,stu84,lam89,tho85}, normalized to the adopted value for $^{106}$Pd.}
 \footnotetext[2]{From \protect \cite{smith05}. The result for $^{114}$Pd corresponds to $\tau=117(6)$ ps \cite{Mach2003,Dewald2008}.}
\end{table}
%

\begin{table}
\caption{$g(2^+_1)$ values in the Cd isotopes.} \label{tab:Cd-data}
\begin{ruledtabular}
\begin{tabular}{cccc}
Nuclide & \multicolumn{2}{c}{$g(2^+)$} \\
 & \multicolumn{1}{c}{\rm previous \footnotemark[1] } & \multicolumn{1}{c}{\rm present and adopted} \\
\hline                & \\
$^{106}$Cd  & 0.40  (0.10)   & +0.393(31) \\
$^{108}$Cd  & 0.34  (0.09)   & +0.389(31) \\
$^{110}$Cd  & 0.285 (0.055) \footnotemark[2] & +0.407(29) \\
$^{112}$Cd  & 0.32  (0.08)   & +0.360(24) \\
$^{114}$Cd  & 0.29  (0.07)   & +0.325(21) \\
$^{116}$Cd  & 0.30  (0.07)   & +0.296(24) \\
\end{tabular}
\end{ruledtabular}
 \footnotetext[1]{From \protect \cite{bre80}.}
 \footnotetext[2]{Calibration value adopted in \protect \cite{bre80}.}
 \end{table}

\section{Tidal wave approach for  $g$-factor calculations}
\label{sect:theory}

As noted in the Introduction, the tidal-wave model uses the fact that in the semi-classical approximation the yrast states of vibrational nuclei correspond to quadrupole waves traveling over the surface of the nucleus, like the tidal waves over the surface of the ocean. In the frame of reference that co-rotates with the wave the quadrupole deformation is static - like that of a rotor. The difference between the vibrator and rotor is that the angular momentum of an ideal rotor is generated by increasing the angular velocity, whereas in the case of the ideal vibrator the angular velocity is constant (equal to half the vibrational frequency) and angular momentum is generated by increasing the
amplitude of the wave. Real nuclei are in between these idealized limits, i.e. both deformation and angular velocity increase with angular momentum. The fact that the yrast states of vibrational, transitional and rotational nuclei correspond, in the semi-classical approximation, to a uniformly rotating statically deformed shape allows the use of the self-consistent cranking model for describing yrast states in transitional nuclei. It was demonstrated in Ref. \cite{frauendgusun1,frauendgusun2} that the energies of the yrast states in even-even nuclei with $44\leq Z \leq 48$ and $54\leq N \leq 68$ are very well described by this model. Individual differences between
the nuclides, which reflect the response of the nucleonic orbits at the Fermi surface to rotation, are  reproduced. The $B(E2)$ values of the transitions between the yrast levels are also well accounted for, including their linear increase with angular momentum
in vibrational nuclei. In this section, the extension of the model to calculate $g$~factors is described.

The details of the tidal wave approach are presented in Refs. \cite{frauendgusun1, frauendgusun2}.
In essence, the self-consistent cranking model is applied to nuclei that are spherical or weakly deformed in their
 ground states.
The calculations are based on the Tilted-Axis-Cranking (TAC) version of the Cranking model as described
in Ref.  \cite{frauendorf00}. 
The cases considered correspond to a rotation of the nucleus about a principal axis.
We start from the rotating mean field
\begin{equation}\label{Hamiltonian}
h^{\prime} =h_{\rm nilsson}(\epsilon,\gamma)+\Delta (P^{+} +P)-\omega J_{x} -\lambda N,
\end{equation}
which consists of quadrupole deformed
Nilsson potential  $h_{\rm nilsson}$ \cite{nilsson95}, combined with a monopole pair field  $P^{+}$
and fixed pair potential $\Delta$. The chemical potential $\lambda$  is fixed for
each deformation such that the particle number is correct for $\omega=0$.

The calculations of the $g$~factors  are performed on a grid of triaxial quadrupole deformations
while the angular frequency $\omega$ is fixed at every grid point by the condition
\begin{eqnarray}
J = \langle\omega(J)| J_x|\omega(J) \rangle+J_c,
\end{eqnarray}
which is facilitated by linear interpolation between discrete $\omega$ grid points.
A small correction is applied to the angular momentum
\begin{equation}\label{eq:Jc}
J_c=100 \; {\rm MeV}^{-1} \varepsilon_2^2\sin^2(\gamma-\pi/3)\omega.
\end{equation}
About half of
it takes into account the coupling between the oscillator
shells  and another half is expected
to come from quadrupole pairing, both being neglected in the Cranking calculations.
For the study of the 2$^+_1$-states $J=2$ was set. The calculations for $J=4$ have also been carried out.
The diabetic tracing technique, as described in Ref.~\cite{frauendorf00},  reliably prevented sudden changes of the
quasiparticle configuration. The tracing was performed by using moderate
steps, $\Delta \omega=0.05$ MeV, and comparing the overlap of configurations step by step.

The total energies  are calculated by means of the  Strutinsky method (SCTAC in \cite{frauendorf00}):
\begin{eqnarray}
E(J,\epsilon ,\gamma )=E_{LD}(\epsilon ,\gamma)
- \tilde{E}(\epsilon ,\gamma)+   \nonumber \\
+\langle\omega(J)| h^{\prime}|\omega(J) \rangle- \langle\omega=0| h^{\prime}|\omega=0 \rangle +\omega(J) J.
\end{eqnarray}
After minimizing the energy $E(J,\epsilon ,\gamma )$  to obtain the
equilibrium deformation parameters,  $\epsilon_e$  and $\gamma_e$,  the magnetic moment is calculated.
\begin{eqnarray}\label{eq:mu}
\mu =   \langle\omega(J),\epsilon_e ,\gamma_e| \mu_{x}|\omega(J),\epsilon_e ,\gamma_e \rangle,
\end{eqnarray}
where
\begin{eqnarray}\label{eq:mux}
\mu_{x} = \mu_{N} (J_{x,p} +(\eta 5.58 -1 ) S_{x,p} - \eta 3.73 S_{x,n} ).
\end{eqnarray}
The spin contributions to the single-particle magnetic moments were evaluated with a common attenuation factor, $\eta=0.7$.
The  possibility that the correction term to the angular momentum, $J_c$, contributes to the magnetic moment
was disregarded.  The $g$~factor is  given by
\begin{eqnarray}\label{eq:gfull}
g(J) = \frac{\mu(J)}{J}.
\end{eqnarray}

The $g$~factors turned out to be sensitive
to the choice of $\Delta_p$ and $\Delta_n$. (See e.g. Ref. \cite{stu95} for a general discussion on the sensitivity of $g$~factors to pairing.)
For the transitional
nuclei around $A=100$, the experimental pairing parameters $\Delta$, as calculated from the
even-odd mass differences, fluctuate considerably with the particle numbers. Using these experimental
values in the calculations translates into  fluctuations of the $g$ factors that are in contradiction to experiment.
The even-odd mass differences do not only reflect the pairing strength but are also
sensitive to the level spacing and deformation changes, which may be the major source  of the fluctuations.
For this reason we adopted constant values of $\Delta$, which are somewhat smaller than the average
experimental values obtained by means of the four-point formula.
The pair gap parameters   $\Delta_p=\Delta_n=1.1$~MeV are adopted for the Mo and Ru isotopes and
$\Delta_p=\Delta_n=$1.2~MeV for the Pd and Cd isotopes. The results for  $\Delta_p=\Delta_n=$1.1~MeV
are also shown for several Pd and Cd isotopes.

In all cases we have reported $g$~factors for the nuclear deformation of minimum calculated energy. For the heaviest isotopes of Mo and Ru the Strutinsky method predicts that the oblate minimum is lowest, with a close by prolate minimum. As with other mean field approaches, the inaccuracies in the prolate-oblate energy difference are of the order of a few hundred keV. In both $^{108}$Mo and $^{112}$Ru there are small energy differences along the gamma degree of freedom. A detailed examination of the dependence of the $g$~factors on the nuclear shape in such cases is beyond the scope of the present report. The effect is expected to be secondary compared to the effects of pairing, which affect the calculated shapes as well as the $g$ factors.

\begin{figure}
  \includegraphics[width=0.45\textwidth]{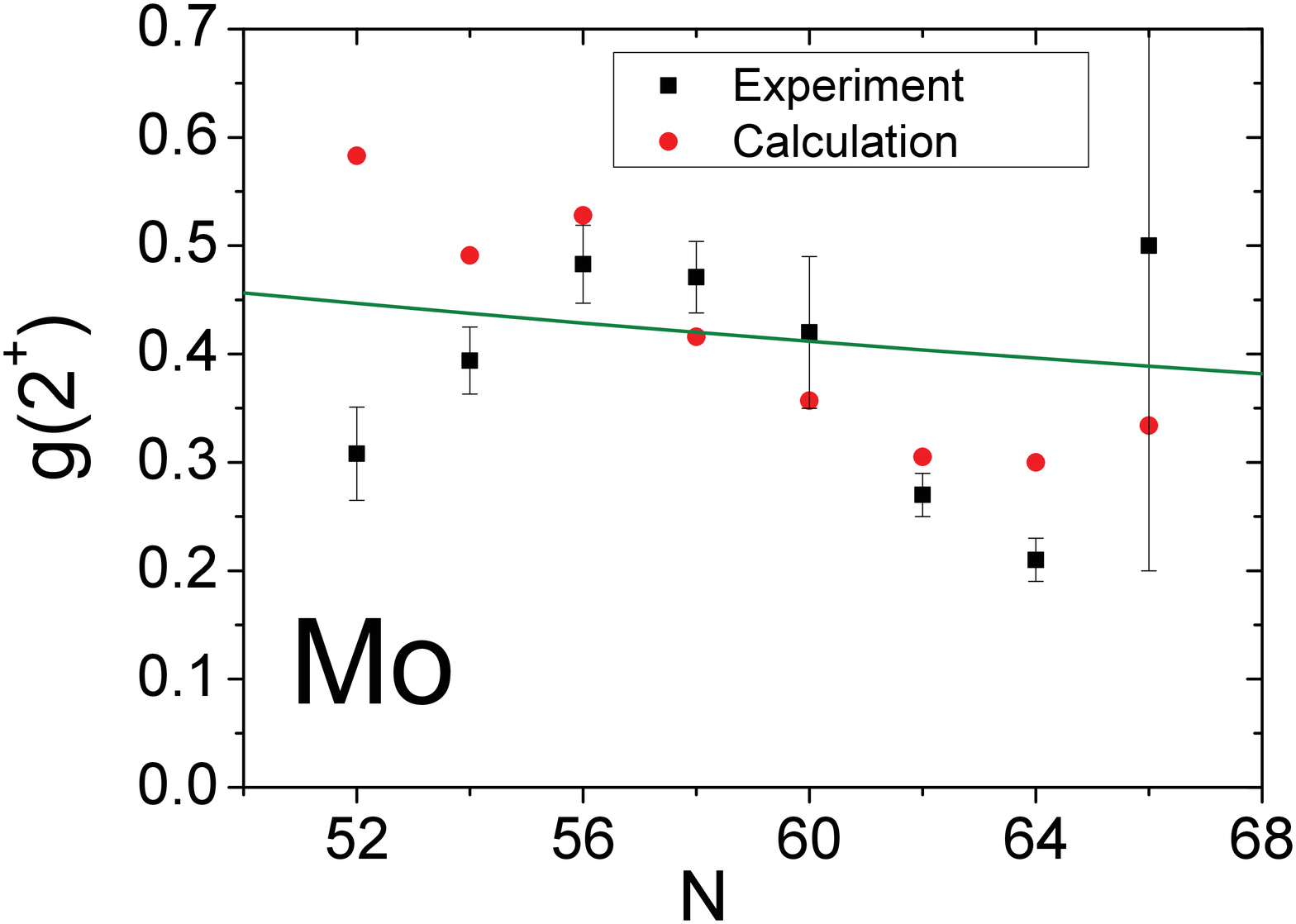}
  \includegraphics[width=0.45\textwidth]{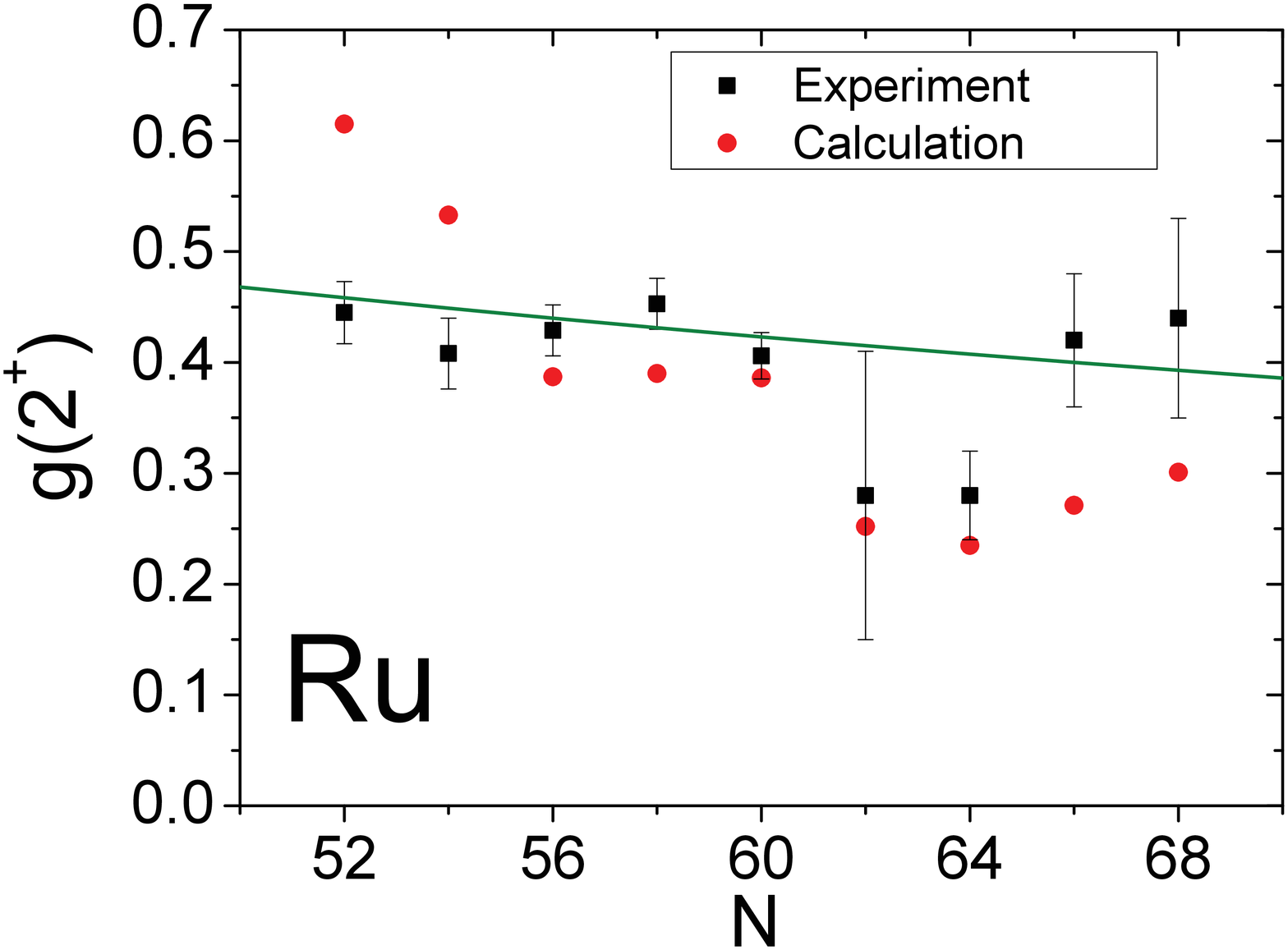}
  \includegraphics[width=0.45\textwidth]{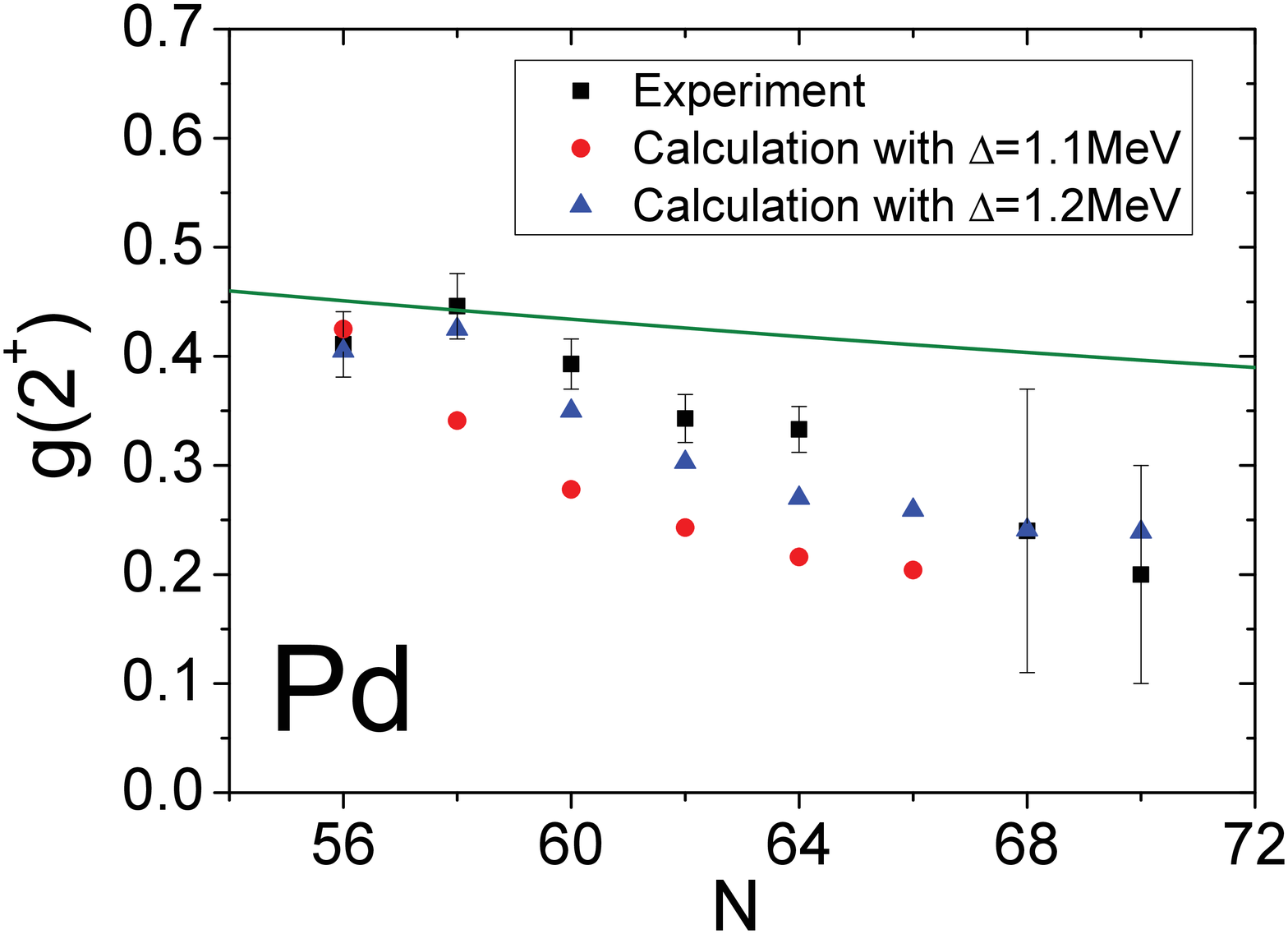}
  \includegraphics[width=0.45\textwidth]{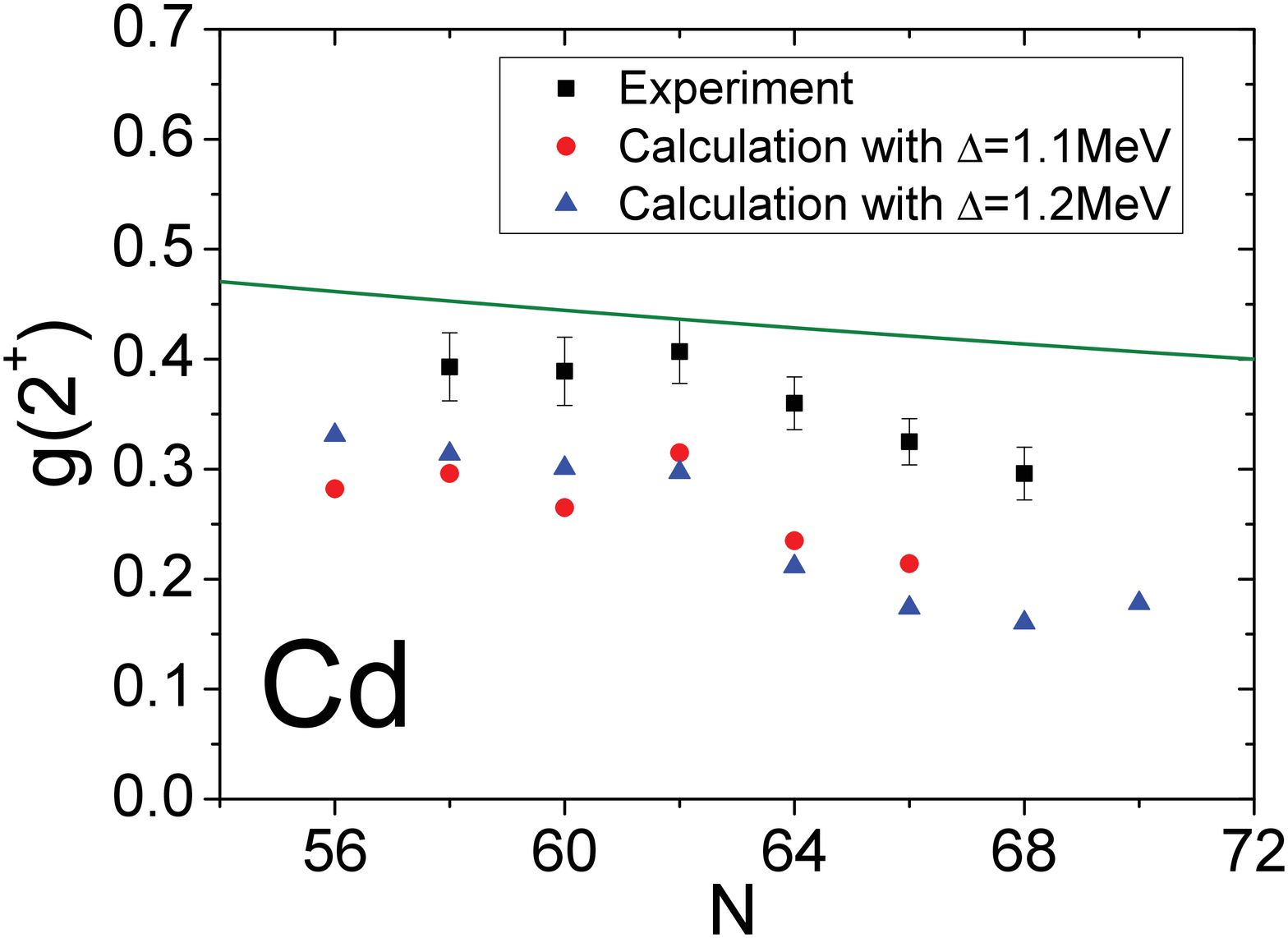}
  \caption{(Color online) $g(2^+)$ calculations compared with experiment. The green line shows $Z/A$.}
\label{fig:ResultFig}
\end{figure}

The results of the calculations are listed in Tab.~\ref{tab:Theory1} and compared with the experimental data
in Fig ~\ref{fig:ResultFig}. By applying the same calculation, the energies, $B(E2)$ values, and $g$~factors, can be obtained
for the yrast sequences in the considered nuclei (see also Refs. \cite{frauendgusun1,frauendgusun2}). The good agreement with experimental data
demonstrates the applicability  of the tidal wave approach.

\begin{table}
\caption{Deformations and calculated $g$~factors.}
\label{tab:Theory1}
\begin{ruledtabular}
\begin{tabular}{ccccccc}
 Nuclide  & \multicolumn{2}{c}{deformation $0^{+}$ }  & \multicolumn{2}{c}{deformation $2^{+}$}  & \multicolumn{1}{c}{$g(2^+)$} & \multicolumn{1}{c}{$g(4^+)$}\\
\cline{2-3} \cline{4-5}
 & $\epsilon_{2}$ & $\gamma$ & $\epsilon_{2}$  &  $\gamma$ &   \\
 \hline							
$_{42}^{92}$Mo$_{50}$	& 0.000	    &   0     & 0.000	    &   0	&   1.323	&        \\
$_{42}^{94}$Mo$_{52}$	& 0.001	    &   5     & $<$0.090    &   0	&   0.583	&        \\
$_{42}^{96}$Mo$_{54}$	& 0.008	    &   5     & 0.117	    &   0	&   0.491	&        \\
$_{42}^{98}$Mo$_{56}$	& 0.169	    &   25    & 0.165	    &   0	&   0.528	&        \\
$_{42}^{100}$Mo$_{58}$	& 0.199	    &   25    & 0.198	    &   0	&   0.416	&        \\
$_{42}^{102}$Mo$_{60}$	& 0.231	    &   25    & 0.239	    &   15	&   0.357	&        \\
$_{42}^{104}$Mo$_{62}$	& 0.252	    &   20    & 0.256	    &   15	&   0.305	&        \\
$_{42}^{106}$Mo$_{64}$	& 0.260	    &   15    & 0.260	    &   15	&   0.300	&        \\
$_{42}^{108}$Mo$_{66}$	& $-0.225$  &   0     & $-0.228$    &   0   &   0.334	&        \\
 							\\
$_{44}^{96}$Ru$_{52}$	& 0.128	    &   0     & 0.130	    &   0	&   0.615	&        \\
$_{44}^{98}$Ru$_{54}$	& 0.119	    &   0     & 0.120	    &   0	&   0.533	&        \\
$_{44}^{100}$Ru$_{56}$	& 0.125	    &   0     & 0.130	    &   0	&   0.550	&   0.468\\
$_{44}^{102}$Ru$_{58}$	& 0.150	    &   0     & 0.151	    &   0	&   0.470	&   0.209\\
$_{44}^{104}$Ru$_{60}$	& 0.175	    &   0     & 0.176	    &   0	&   0.350	&   0.145\\
$_{44}^{106}$Ru$_{62}$	& 0.200	    &   0     & 0.200	    &   0	&   0.324	&   0.194\\
$_{44}^{108}$Ru$_{64}$	& 0.210	    &   0     & 0.210	    &   0	&   0.310	&   0.193\\
$_{44}^{110}$Ru$_{66}$	& 0.210	    &   0     & 0.211	    &   0	&   0.350	&   0.231\\
$_{44}^{112}$Ru$_{68}$	& $-0.210$  &   0     & $-0.210$    &   0	&   0.301	&        \\
						\\
$_{46}^{102}$Pd$_{56}$	& 0.100	    &   0     & 0.120	    &   0	&   0.405	&   0.611\\
$_{46}^{104}$Pd$_{58}$	& 0.125	    &   0     & 0.135	    &   0	&   0.425	&   0.504\\
$_{46}^{106}$Pd$_{60}$	& 0.126	    &   0     & 0.150	    &   0	&   0.350	&   0.320\\
$_{46}^{108}$Pd$_{62}$	& 0.150	    &   0     & 0.151	    &   0	&   0.303	&   0.204\\
$_{46}^{110}$Pd$_{64}$	& 0.152	    &   0     & 0.154	    &   0	&   0.270	&   0.186\\
$_{46}^{112}$Pd$_{66}$	& 0.154	    &   0     & 0.156	    &   0	&   0.259	&   0.222\\
$_{46}^{114}$Pd$_{68}$	& 0.190	    &   0     & 0.191	    &   0	&   0.241	&        \\
$_{46}^{116}$Pd$_{70}$	& 0.174	    &   0     & 0.170	    &   0	&   0.239	&        \\
								\\
$_{48}^{104}$Cd$_{56}$	& 0.000	    &   0     & 0.090	    &   0	&   0.331	&   0.288\\
$_{48}^{106}$Cd$_{58}$	& 0.050	    &   0     & 0.100	    &   0	&   0.314	&   0.327\\
$_{48}^{108}$Cd$_{60}$	& 0.075	    &   0     & 0.114	    &   0	&   0.302	&   0.163\\
$_{48}^{110}$Cd$_{62}$	& 0.050	    &   0     & 0.100       &   0	&   0.297	&   0.142\\
$_{48}^{112}$Cd$_{64}$	& 0.000	    &   0     & 0.120	    &   0	&   0.212	&   0.125\\
$_{48}^{114}$Cd$_{66}$	& 0.050	    &   0     & 0.117	    &   0	&   0.174	&   0.117\\
$_{48}^{116}$Cd$_{68}$	& 0.070     &   0     & 0.090	    &   0 	&   0.161	&        \\
$_{48}^{118}$Cd$_{70}$	& 0.100     &   0     & 0.110	    &   0 	&   0.178	&        \\

\end{tabular}
\end{ruledtabular}
\end{table}

\section{Discussion}
\label{sect:discussion}

\subsection{$g$ factor trends in the Tidal Wave Model}

The $N$-dependence of $g$~factors in transitional nuclei has been a challenge to theory.
The main reason is that the $g$ factors are sensitive to the underlying single particle composition of the collective
quadrupole degree of freedom. The collective states of transitional nuclei have been mostly described in the frame work
of phenomenological collective models such as the Bohr Hamiltonian \cite{BMvol2p532} and the Interacting Boson Model \cite{IAp42}, which do not specify the fermionic structure of the collective mode. On this level, one simply
assumes that only the protons are responsible for the the current that generates the magnetic moment, i.e. $g=Z/A$.
Sambataro and Dieperink  \cite{Sambataro1981} addressed the experimental deviations of the $g$~factors of the transitional isotopes of Ru to Te in the framework of proton-neutron Interacting Boson Model. Since then,  there has been little progress towards a deeper microscopic-based interpretation.
 We approach the problem anew within the tidal wave model \cite{frauendgusun1,frauendgusun2}, which
 is completely microscopic. It describes the yrast states  by means of the self-consistent Cranking model,
which allows one to calculate the magnetic moment directly from the nucleonic currents. In applying it to
nuclei that are spherical or slightly deformed in their ground state, one has to numerically diagonalize the
quasiparticle Routhian, Eq.~(\ref{Hamiltonian}). Only for well deformed nuclei is the perturbative expression,
as given in Ref.~\cite{PriorBoehmNilsson68}, applicable.
The calculations in Refs. \cite{frauendgusun1,frauendgusun2} indicate that
the nuclides in the considered  transitional region are anharmonic  vibrators or very soft rotors.
 The deformation of the $2^+$ state  increases with the number of valence proton holes below, and neutron particles above, the $Z=N=50$ shell.
It ranges from $\epsilon_2\sim0.1$ for the vibrational Cd isotopes to $\sim 0.25$ for the rotational nuclei $^{102,104,106}$Mo.
 For each isotope chain, the deformation increases with the neutron number.
Fig.~\ref{fig:ResultFig} shows that the calculated $g$~factor trends are in overall good agreement with experiment.
In particular, the deviations from the value $Z/A$ are well accounted for.

\begin{figure}
 \includegraphics[width=0.45\textwidth]{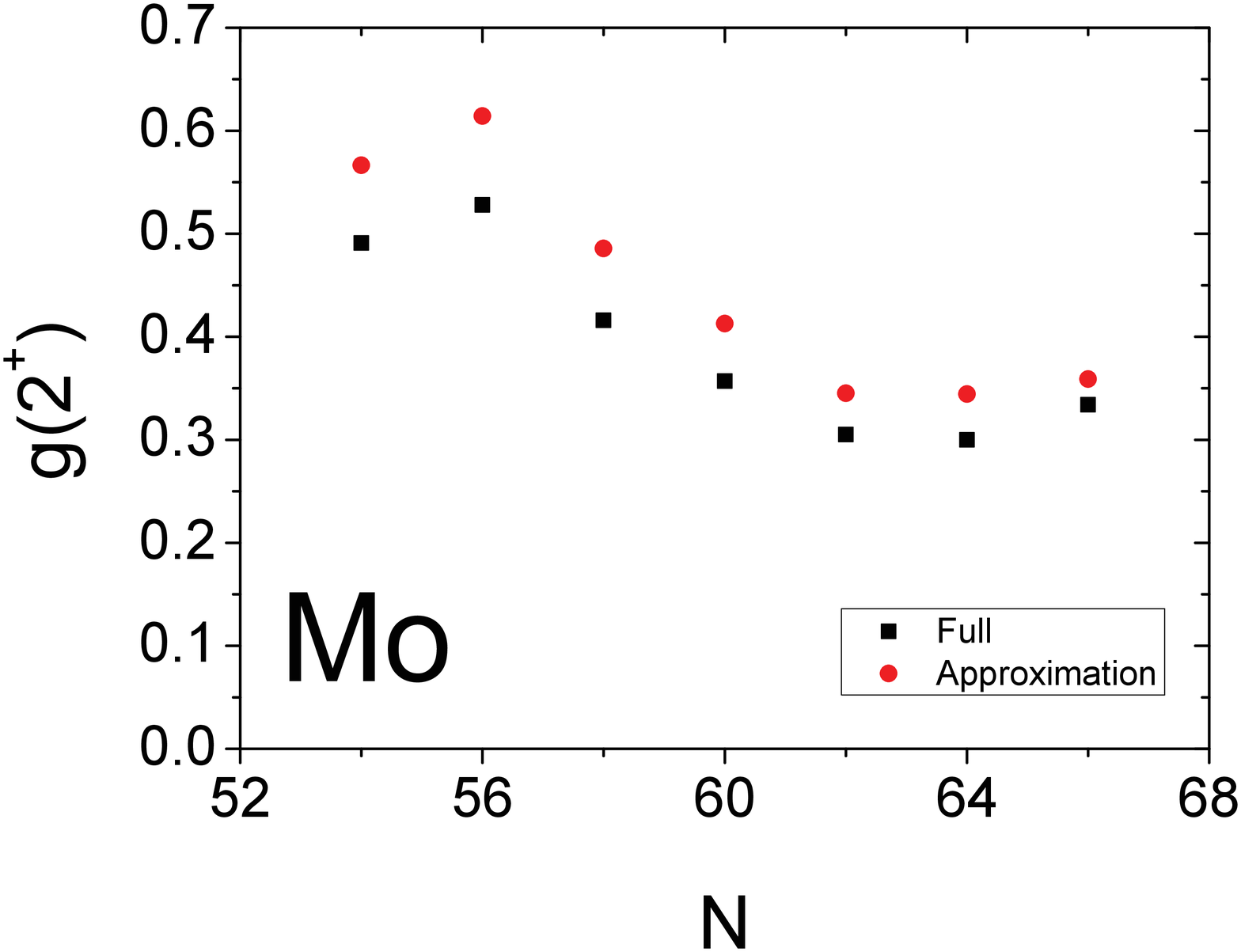}
 \includegraphics[width=0.45\textwidth]{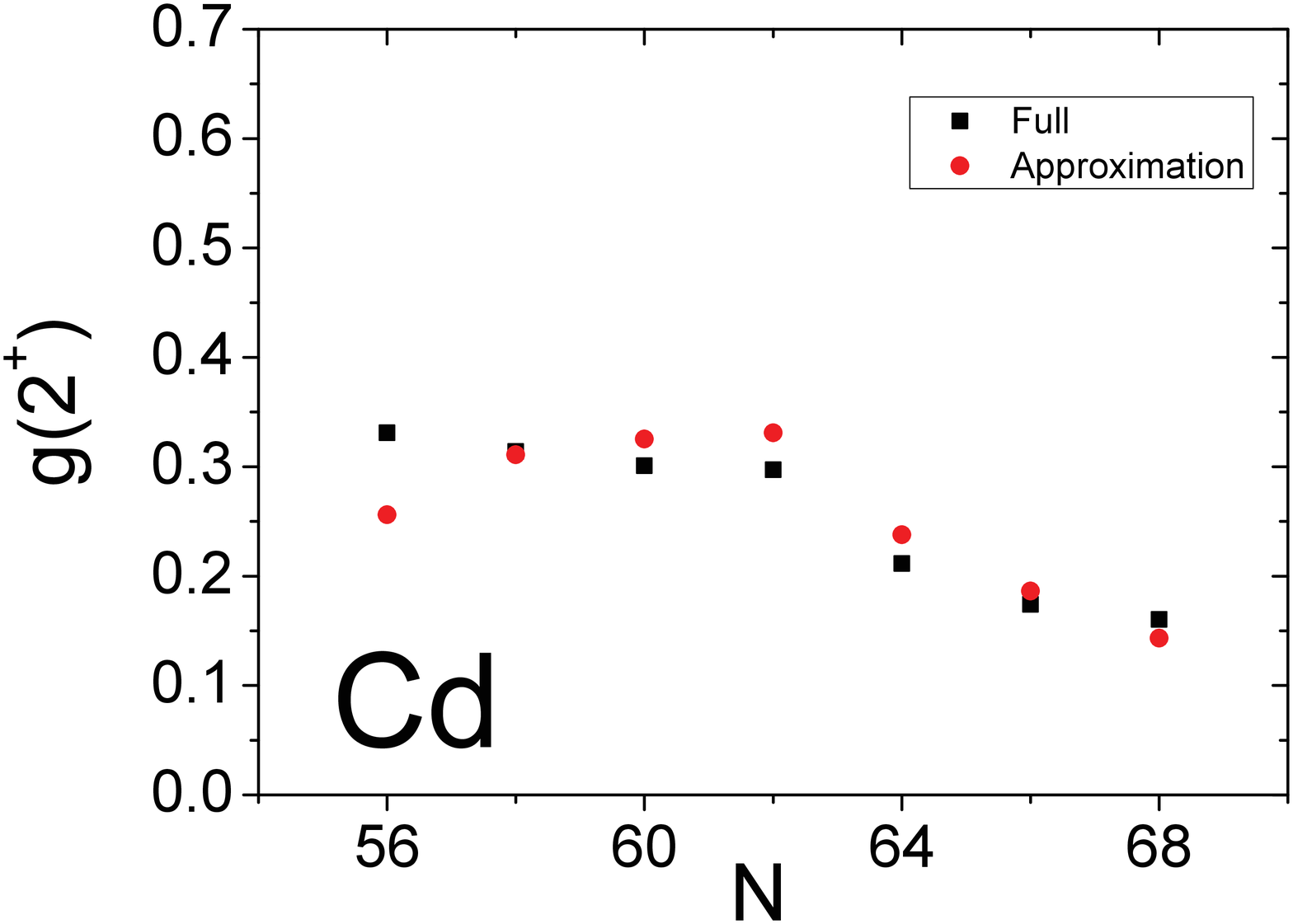}
\caption{(Color online) Comparison between calculated $g$~factors.
  Black squares show the full calculation as described in the text and shown in Fig.~\ref{fig:ResultFig}. Red dots show the approximation of Eq.~(\ref{eq:gapprox}), which takes into account only the magnetic moments generated by the
  $g_{9/2}$ protons and the $h_{11/2}$ neutrons.}
\label{fig:gapprox}
\end{figure}

The $Z$ - and $N$ - dependence of the $g$~factors will now be discussed.
Looking at the microscopic contributions to the angular momentum, it is seen that a few quasiparticle levels near the Fermi surface contribute most to the total angular momentum.
The $g$~factors generally decrease along the isotope chains. This decrease is the result of the fractional increase of $J_{n5}$, the angular momentum contribution from the $h_{11/2}$ neutrons.
To elucidate this observation,  Fig.  \ref{fig:gapprox}
compares the calculated $g$ factors  with the simple approximation
\begin{eqnarray}\label{eq:gapprox}
g(J)\approx \left(1.43 J_{p4}-0.24J_{n5}\right)/2,
\end{eqnarray}
where $J_{p4}$ and $J_{n5}$ are the angular momenta calculated by means of the Cranking model
for the proton $N=4$ and the neutron $N=5$ shells, respectively.
This approximation  assumes that (i) only the $g_{9/2}$ proton holes and the $h_{11/2}$ neutrons contribute to the
magnetic moment, and (ii) the expression
\begin{equation}\label{eq:gsph}
g=g_l\pm\eta \frac{g_s}{2l+1},
\end{equation}
which is valid for the spherical shells, can be used to set the nucleon $g$~factors. The contribution from the neutron $N=4$ shell
is assumed to be zero because  it is generated by the $d_{3/2}$ and $d_{5/2}$ orbitals, which have opposite
$g$~factors.   Fig. \ref{fig:gapprox} compares the approximation of Eq.~(\ref{eq:gapprox}) with the full calculation.
 It is seen that the approximation qualitatively accounts for the $N$ dependence of the $g$ factors.
 The separate contributions of the various oscillator shells, $N$,  to the
 total magnetic moment were calculated  by means of  Eq.~(\ref{eq:mux}). It turned out that the contributions of the $N=3$ proton and neutron shells, and of the
 $N=4$ neutron shell, are negligible. The errors of the approximation in Eq.~(\ref{eq:gapprox}) are therefore due to the use of
 the single-nucleon $g$~factors for  spherical $j$-shells. Nevertheless, the simplified expression, Eq.~(\ref{eq:gapprox}), allows one to understand the $N$ and $Z$ dependence of the excited-state $g$ factors.

\begin{figure}[h]
 \includegraphics[width=0.45\textwidth]{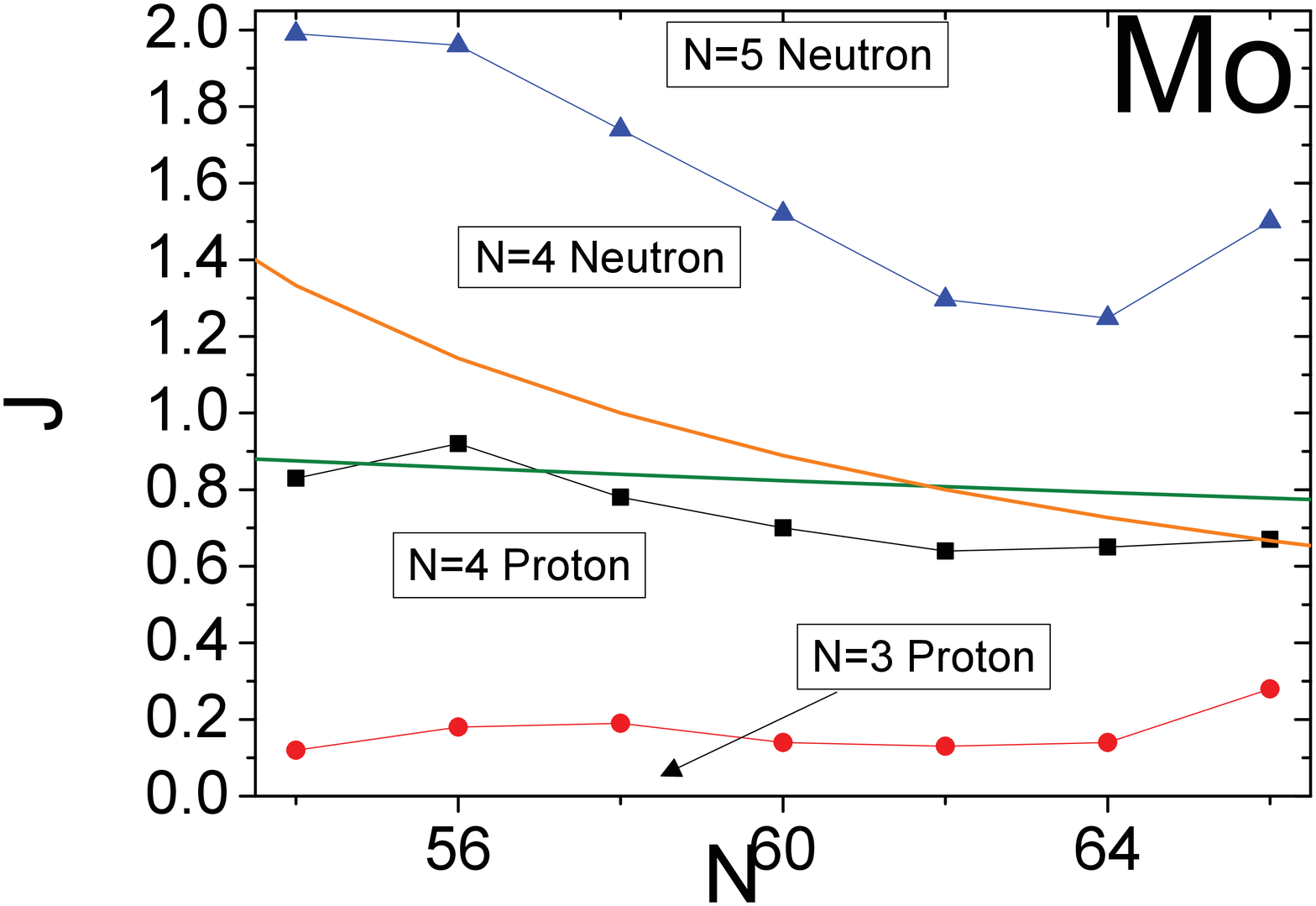}
  \includegraphics[width=0.45\textwidth]{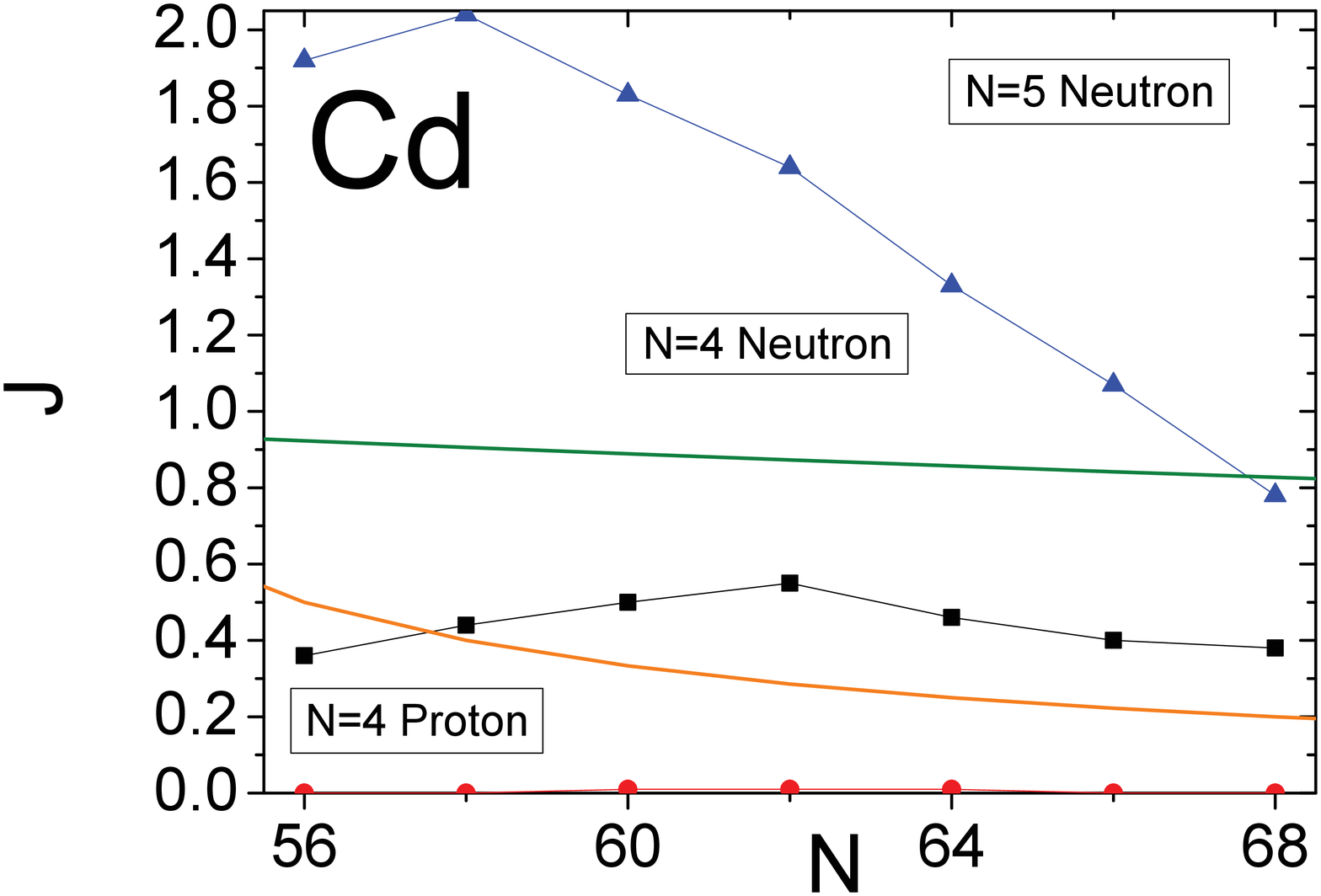}
 \caption{(Color online) Composition of the angular momentum of the 2$^+$ states of the
  Mo and Cd isotopes. The fraction of $J=2$ of each oscillator shell as obtained from the
  Cranking model is shown. The distance between the lower frame and the circles (red curve) is the fraction of the protons in the $N=3$ oscillator shell,
  the distance between the circles and squares (red and black curves)  the $N=4$ proton fraction, the distance between the squares and triangles (black and blue
  curves) the $N=4$ neutron fraction, and the distance between the triangles (blue curve) and the upper frame the $N=5$ neutron fraction. The contribution of the $N=5$ protons is negligible.
 The distance between the lower frame and orange curve with no symbols is the
  proton fraction according to
   the IBM-II boson counting rule calculated by means of Eq.~(\ref{eq:JpIBA}).  The neutron fraction is the distance between
   this curve and the upper frame. The straight green line shows $2Z/A$.}
   \label{fig:JpJn}
\end{figure}

\begin{figure}
  \includegraphics[angle=0,width=0.47\textwidth]{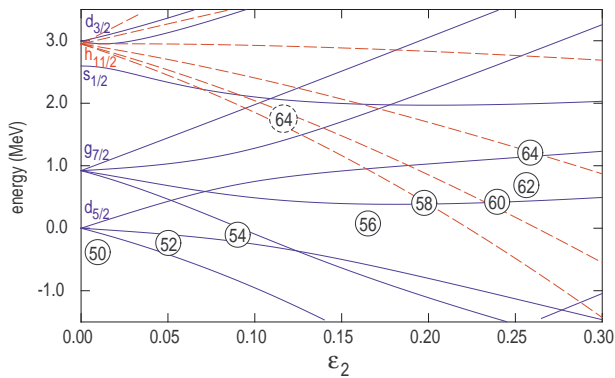}
  \caption{(Color online) Nilsson diagram of Molybdenum neutrons close to the $N=50$ shell, the dash lines represent negative parity and belong to $h_{11/2}$ levels. The neutron numbers of Mo isotopes are shown at the calculated equilibrium deformation, which illustrates the intrusion of the
  $h_{11/2}$ shell among the positive parity orbits for $N < 64$, and explains the impact of $h_{11/2}$ orbit on the calculated $g$~factors.
   }
\label{fig:Moneutron}
\end{figure}

As an example, the behavior of the  Mo isotopes was examined in greater detail. Fig. \ref{fig:JpJn} shows the composition of the angular momentum of the 2$^+$ state. With increasing numbers of valence neutrons, the $g_{9/2}$ proton fraction,  $J_{p4}$, remains nearly constant for the isotope chain, whereas the $h_{11/2}$ neutron fraction, $J_{n5}$, increases, which causes  the decrease of the $g$~factor.
As illustrated in Fig.~\ref{fig:Moneutron}, the increased neutron numbers  push up
the Fermi surface, and with increased deformations the Fermi level moves into the lower half of the Nilsson orbits with $h_{11/2}$ parentage, which generate $J_{n5}$. Since the $h_{11/2}$ neutrons have $g=-0.24$, they progressively reduce
the magnetic moment. In the Cd isotopes the proton fraction,  $J_{p4}$, is smaller because there are only two $g_{9/2}$
holes. The progressive occupation of the $h_{11/2}$ neutron orbitals  generates the $J_{n5}$ fraction seen in Fig. \ref{fig:JpJn},
which causes the decrease of the $g$~factor clearly seen in Fig. \ref{fig:gapprox} for the simplified expression, Eq.~(\ref{eq:gapprox}).

For the Cd isotopes, the deformation of the 2$^+$ state is about 0.1. As seen in Fig.  \ref{fig:Moneutron},
the Fermi level reaches the $h_{11/2}$ states only at $N=64$.
 However one has to keep in mind that the smaller
deformation implies a higher angular frequency of the tidal wave, which lowers the $h_{11/2}$ orbitals relative to
the positive parity orbitals. For this reason, the occupation of the $h_{11/2}$ orbitals starts already at $N=60$.
Hence, in the studied region, the neutrons in the $h_{11/2}$ orbit are primarily responsible  for the drop of the $g$ factors with increasing neutron number.
This observation agrees with the inference of Smith {\em et al.} \cite{smith04} based on their measured $g$~factors for several neutron-rich isotopes. Moreover,  it has long been known that  the strong increase of  $J_n$ in well deformed nuclei, caused by the rotational alignment
of the $i_{13/2}$ and $j_{15/2}$ neutrons, reduces the  $g$ factors below $Z/A$ \cite{ChenFrau83}.

The  $g$ factors of the Cd isotopes are  underestimated by about 20\%. Reducing the proton pair gap $\Delta_p$ by
about 10\%, while keeping the neutron gap $\Delta_n$ unchanged, would increase the proton fraction, $J_p$, relative to
the neutron fraction, $J_n$, such that  the  $g$ factors have the correct magnitude. The $N$ dependent trend will not be
changed. A reduced  $\Delta_p$ for $Z=48$ appears reasonable, because there are only two proton holes to
generate pair correlations. The other isotopes have more proton holes, which generate stronger the pair correlations.

The ``canonical" estimate  $g=Z/A$ for collective quadrupole excitations is based on the assumptions that (i) the ratio
$J_p/(J_p+J_n)=Z/A$ and (ii) that the spin contributions of the protons and neutrons cancel.  Assumption (i) is rather
poor for the Cd isotopes, which are almost semi magic. It becomes better for Mo isotopes, which are situated further
into the open shells. Assumption (ii) is not justified for the high-spin intruder orbitals $g_{9/2}$ and $h_{11/2}$, which almost completely generate the magnetic moment. Although (i) and (ii) become more valid assumptions for increasing numbers of valence nucleons,
the differences between  the $g$~factors of the nucleonic orbitals near the Fermi surface remain noticeable in the $Z$ and $N$
dependence of nuclear $g$~factors (cf. Ref.~\cite{ChenFrau83}).

\subsection{$N=50, 52$ cases: $^{92,94}$Mo, $^{96}$Ru}

The tidal wave approach does not work for  the $N=50$ spherical nucleus $^{\ 92 }$Mo.
The ground state configuration is not able to
generate enough angular momentum to reach $J=2$. By recognizing the fact that the neutrons
of $^{\ 92 }$Mo complete a  shell, the lowest  $2^+$ configuration is
 obtained as the two quasi-proton configuration with angular momentum projection of 2.  Its
 deformation is found to be zero. The calculated $g$~factor of 1.32 is close to the value 1.43 for
 the spherical $g_{9/2}$ proton orbital. The $N=52$  nuclide $^{94}$Mo also turns out to be not quite amenable  to the tidal wave
 approach. Assuming a deformation of $\epsilon_2=0.09$, one can generate 2 units of angular momentum at
 a frequency that is consistent with the energy of the 2$^+$ state. However, this deformation is not stable and the equilibrium deformation lies at a smaller value.   As seen in Fig. \ref{fig:JpJn}, $J_{p4}$=0.8, which generates
 the magnetic moment. The neutron fraction, $J_{n4}=1.2$, contributes very little due to a cancelation of the single-neutron components, particularly the $d_{5/2}$ and $g_{7/2}$ orbits (see Fig.~\ref{fig:Moneutron}) which have single-particle moments of opposite sign. The resulting
 $g$~factor of 0.58 is much larger than the experimental value of $0.31\pm0.04$.

 Recently Holt {\em et al.} \cite{ Holt07} carried out shell model calculations, finding $g\approx 0.2$. They state  that  $J_{n4}$ originates mainly from the $d_{5/2}$ orbital. Assuming
 that  our value of $J_{n4}=1.2$ exclusively originates from  the $d_{5/2}$ orbital and using $g(d_{5/2})=-0.52$ from Eq.~(\ref{eq:gsph}),
 one finds $g(2^+)=(1.43\times0.8-0.52\times1.2)/2=0.26$, which is close to experiment.
 As compared with the shell model, in our calculations only pairing and quadrupole correlations are taken into
 account. It seems that the quadrupole correlations are over estimated, which causes an increased $g$~factor through admixtures of the $g_{7/2}$
 orbital.

 In the case of the $N$=52 nuclide $^{96}$Ru, the tidal wave approach gives a finite value of $\epsilon_2=0.13$ and a good estimate for the frequency of the 2$^+$ state. The calculated $g$~factor is too large for the same reason as for $^{94}$Mo, however the discrepancy is less, which is likely a consequence of the increasing configuration mixing with the number of valence nucleons and deformation.

\subsection{$g$~factors of the 4$^+$ states}

The tidal wave model can predict the $g$~factors of higher excited yrast states above the 2$^+$ state, and a number of such predictions have been included in Table \ref{tab:Theory1}. Unfortunately the experiments to measure these $g$~factors are challenging and experimental data are scarce. To our knowledge the only measurement to date is very recent work by G\"urdal {\em et al.} \cite{Gurdal2010} on $^{106}$Pd, where $g(4^+)=0.44\pm 0.09$ was obtained, relative to $g(2^+)=0.393\pm 0.023$. Within uncertainties, the experimental $g$-factor ratio is consistent with the present predictions. Similar experiments on the heavier isotopes such as $^{110}$Pd could decisively detect if  $g(4^+) < 0.5 g(2^+)$ as predicted by the tidal wave model calculations.

\subsection{Comparison with Interacting Boson Model}

The present study gives new insights into the relative angular momentum carried by protons and neutrons in the transitional nuclei near $A=100$, which can be compared and contrasted with the detailed analysis of Sambataro and Dieperink  \cite{Sambataro1981} based on the interacting boson model.
These workers studied the $g$~factors in this region in the framework
of IBM-II, which distinguishes between proton and neutron bosons.  The model parameters were fitted to the
energies and $B(E2)$ values of the lowest collective quadrupole excitations. They found that, to a good approximation,
$ J_p \propto N_p$ and $J_n \propto N_n$,
where $N_p$ and $N_n$ are the number of proton and neutron bosons, respectively. According to the IBM-II counting
rule, these numbers are equal to one half of the number of valence proton holes and one half of the number of valence neutrons relative to $Z=N=50$,
respectively, i.e. for $J=J_p+J_n=2$ one has
\begin{equation}\label{eq:JpIBA}
J_p=2\frac {N_p}{N_p+ N_n}=2\frac {50-Z}{50-Z+ N-50}=2\frac {50-Z}{N-Z}.
\end{equation}

Sambataro and Dieperink assigned effective $g$~factors to the
proton and neutron valence systems, $g_p(N_p)$ and  $g_n(N_n)$, which were assumed to depend on $N_p$ and $N_n$.
Considering $g_p(N_p)$ and  $g_n(N_n)$ as free parameters, they fitted experimental $g$ factors. They found that the resulting   $g_p(N_p)$ and  $g_n(N_n)$ values change smoothly with the boson numbers, and claimed that they qualitatively correlate with the $g$ factors of the valence neutrons and protons that constitute the collective quadrupole mode.

Fig. \ref{fig:JpJn} compares the IBM-II values of $J_p$  and $J_n$ with our values, which are also are generated by
the valence particles and holes. As seen, the IBM-II does not track closely with
the proton-neutron ratio of the angular momentum obtained from our microscopic calculation.
More important, it does not provide any information about the composition of the proton and neutron fractions,
which is decisive for the calculation of the magnetic moments.
These deficiencies are overcome by introducing effective $N_p$- and
$N_n$-dependent boson $g$~factors, which are then adjusted to the experiment.
In contrast, our discussion above demonstrates that the $N$- and $Z$ - dependence of the measured $g$ factors is well understood
in terms of the microscopic $J_p$ and $J_n$ fractions and {\it constant}  $g$ factors for the $g_{9/2}$ proton holes
and $h_{11/2}$ neutrons.

\section{Conclusion} \label{sect:conclusion}

The $g$~factors of the first excited 2$^+$ states in all of the stable even isotopes of Mo, Ru, Pd and Cd have been studied experimentally and theoretically. An extensive set of measurements, using variations of the transient-field technique, was completed to ensure that the data set is internally consistent, i.e. the relative $g(2^+_1)$ values are accurate both within and between the isotope chains. Absolute values of the $g$ factors were set relative to $g(2^+_1)$ in $^{106}$Pd. The experimental precision has been improved considerably.

The data have been compared in detail with the tidal wave version of the cranking model. We conclude that the tidal wave approach gives a convincing description of the mass-dependent $g$-factor systematics in vibrational and transitional nuclei. Moreover the $g$ factors reveal the proton-neutron composition of the collective quadrupole mode. In comparison with previous work in this region based on the proton-neutron interacting boson model (IBM-II), the tidal wave model is more solidly based on the underlying single-particle structure. It is found that the simple IBM-II counting rule based on the valence proton fraction gives only a rough guide. In particular, the individuality of the valence nucleons (especially their single-particle $g$~factors) must be considered explicitly.

Looking to the future, on the experimental side it is feasible to test the spin-dependent predictions of the tidal-wave model in a number of cases. In terms of improving the theory, it has been noted that the $g$~factors are very sensitive to the relative strength of neutron and proton pairing. More accurate predictions than those presented here will require a more sophisticated, self consistent treatment of pairing.

\begin{acknowledgments}
The authors are grateful to the academic and technical staff of the
Department of Nuclear Physics (Australian National University) for
their assistance. Alan Devlin is thanked for contributions to the data collection. Mike Taylor,  No\'emie Koller and Gerfried Kumbartzki are thanked for making available Target III and for discussions. This work was supported in part by the Australian
Research Council Discovery Scheme Grant No. DP0773273. MCE acknowledges support from the Australian Postgraduate Award (APA) scheme. SF and JS acknowledge support by the DoE Grant DE-FG02-95ER4093.
\end{acknowledgments}

\bibliography{TF}

\end{document}